\documentclass[10pt]{article}

\usepackage{amsmath}
\usepackage{amssymb}
\usepackage[colorlinks, bookmarksopen, 
             linkcolor={blue},
             anchorcolor={black},
             citecolor={blue},
             filecolor={magenta},
             menucolor={blue},
             pagecolor={red},
             plainpages=false,pdfpagelabels,
             urlcolor={black}]{hyperref}
\usepackage[usenames]{color}
\usepackage{subfigure}
\usepackage{listings}
\usepackage{algpseudocode}
\usepackage{algorithm}
\usepackage{tabulary}
\usepackage{booktabs}
\usepackage{bm}
\usepackage{graphicx}
\algrenewcommand{\algorithmiccomment}[1]{\hskip3em // #1}
\definecolor{lightlightgray}{gray}{0.97}
\definecolor{lightlightgreen}{rgb}{0.8,1,0.8}
\definecolor{darkblue}{rgb}{0,0,0.8}
\definecolor{darkgreen}{rgb}{0,0.5,0}
\newcommand{\Eqref}[1]{Eq.~\ref{#1}}

\newcommand{\Secref}[1]{Sect.~\ref{#1}}

\newcommand{\Algreff}[1]{Algorithm~\ref{#1}}
\newcommand{\Figref}[1]{Fig.~\ref{#1}}
\newcommand{\Figrefs}[1]{Figs.~\ref{#1}}

\newcommand{\Tableref}[1]{Table~\ref{#1}}

\newcommand{\R}{\mathbb{R}}
\newcommand{\KL}{Karhunen-Lo\'{e}ve\ }
\title{Embedded Ensemble Propagation for Improving Performance, Portability and Scalability of Uncertainty Quantification on Emerging Computational Architectures\footnote{This material is based upon work supported by the U.S. Department of Energy, Office of Science, Office of Advanced Scientific Computing Research (ASCR) as well as the National Nuclear Security Administration, Advanced Technology Development and Mitigation program.  This research used resources of the Oak Ridge Leadership Computing Facility, which is a DOE Office of Science User Facility.}}

\author{
E. Phipps, M. D'Elia, H. C. Edwards, M. Hoemmen, J. Hu, S. Rajamanickam \\
Center for Computing Research, \\
Sandia National Laboratories,\footnote{Sandia National Laboratories is a multi-program laboratory managed and operated by Sandia Corporation, a wholly owned subsidiary of Lockheed Martin Corporation, for the U.S. Department of Energy's National Nuclear Security Administration under contract DE-AC04-94AL85000. } \\
Albuquerque, NM and Livermore, CA\\
\url{etphipp, mdelia, hcedwar, mhoemme, jhu, srajama@sandia.gov}}

\begin{document}
\maketitle

\renewcommand{\thefootnote}{\fnsymbol{footnote}}

\renewcommand{\thefootnote}{\arabic{footnote}}

\begin{abstract}
Quantifying simulation uncertainties is a critical component of rigorous predictive simulation.  A key component of this is forward propagation of uncertainties in simulation input data to output quantities of interest.  Typical approaches involve repeated sampling of the simulation over the uncertain input data, and can require numerous samples when accurately propagating uncertainties from large numbers of sources.  Often simulation processes from sample to sample are similar and much of the data generated from each sample evaluation could be reused.  We explore a new method for implementing sampling methods that simultaneously propagates groups of samples together in an embedded fashion, which we call embedded ensemble propagation.  We show how this approach takes advantage of properties of modern computer architectures to improve performance by enabling reuse between samples, reducing memory bandwidth requirements, improving memory access patterns, improving opportunities for fine-grained parallelization, and reducing communication costs.  We describe a software technique for implementing embedded ensemble propagation based on the use of C++ templates and describe its integration with various scientific computing libraries within Trilinos.  We demonstrate improved performance, portability and scalability for the approach applied to the simulation of partial differential equations on a variety of CPU, GPU, and accelerator architectures, including up to 131,072 cores on a Cray XK7 (Titan).
\end{abstract}



\pagestyle{myheadings}
\thispagestyle{plain}



\section{Introduction}
\label{sec:intro}
As computing power continues to increase there is an increasing desire to leverage computational simulation to predict physical, biological, and other scientific phenomena.  However for predictions based on simulation to be rigorously justified, all errors arising in the simulation must be accurately accounted for.  All simulations involve input data, such as physical properties, initial conditions, boundary conditions, and geometries, that are rarely known precisely.  Characterizing the error, or uncertainty, in these data, propagating this uncertainty to the simulation results, and understanding its implications on predictions is generally called \emph{uncertainty quantification}.  Often uncertainty in input data is represented through bounds on the data or random variables/stochastic processes with prescribed probability distribution functions.  Numerous approaches for propagating this uncertainty through computational simulations have been investigated in the literature, including random sampling~\cite{Fishman_96,Helton_Davis_03,McKay_79,Metropolis_49,Niederreiter:1978wr}, stochastic collocation~\cite{Babuska:2007kv,Nobile:2008kv,Nobile:2008dr,Xiu:2005ia}, and stochastic Galerkin~\cite{Ghanem:1990p7167,Ghanem_Spanos_91,Xiu:2002p919}, many of which involve sampling the simulation over the range of uncertain data.  While the accuracy and scalability properties of numerical uncertainty propagation schemes vary, they all suffer from the basic challenge that accurately resolving discontinuous and localized behavior over high-dimension uncertain input spaces can require a tremendous number of samples.  Since each sample evaluation is independent, it is trivial to parallelize sampling-based uncertainty propagation schemes in a coarse-grained manner by executing each sample on a disjoint set of compute nodes (and evaluation of the simulation at each sample typically involves the bulk of the computational cost). However, most uncertainty quantification problems of scientific and engineering interest involve high-fidelity simulations that require for each sample a significant fraction of the available computational resources, and therefore it is often possible to execute only a small fraction of the total number of samples needed in parallel with the remaining fraction executed sequentially.   This leads to very high computational cost, making many uncertainty quantification problems applied to high-fidelity simulations intractable.

In this work, we investigate whether the aggregate computational cost for propagating uncertainties can be reduced by
propagating multiple samples, which we call \emph{ensembles}, together through simulations.  We discuss an embedded approach
for propagating ensembles, whereby scalars within the simulation code are replaced by ensemble arrays, and apply the
technique to the solution of partial differential equations on unstructured meshes.  We demonstrate that the approach, \emph{embedded ensemble propagation}, results in improved performance, portability, and scalability on contemporary multicore and manycore computational architectures by sharing sample-independent data and calculations across the ensemble (reducing floating-point operations and/or memory bandwidth requirements), improving memory access patterns by replacing sparse gather/scatter operations with packed loads/stores, improving opportunities for fine-grained vector (SIMD/SIMT) parallelization, and reducing aggregate communication costs arising from latency.  

The paper is organized as follows.  We discuss the mathematical formulation of ensemble propagation and its impact on iterative linear solvers in \Secref{sec:ensemble}.  In \Secref{sec:software}, we describe a software implementation of the technique leveraging C++ templates and operator overloading, focusing on its integration with the performance portability package Kokkos~\cite{Kokkos:2012:SciProg,Kokkos:2014:JPDC} and with sparse linear algebra tools available in Trilinos~\cite{TrilinosTOMS,TrilinosSP}. While the performance portability of Kokkos has been demonstrated in small mini-applications and graph algorithms~\cite{Kokkos:2012:SciProg,Kokkos:2014:JPDC,KokkosGraph:2015:IPDPS}, \Secref{sec:software} details how the ensemble propagation approach is embedded into the Trilinos solver stack, from linear algebra kernels to iterative solvers and multigrid preconditioners.  Then in \Secref{sec:results} we demonstrate performance improvements for the  approach compared to traditional one-at-a-time sample propagation on a variety of contemporary computational architectures, including weak-scaling results on a Cray XK7 (Titan) with up to 131,072 processor cores.  Finally in \Secref{sec:conclusions} we summarize our work.  

The idea of propagating multiple samples together has been proposed previously~\cite{Giering:2012}.  However, in this work we describe a unique approach for implementing the technique in a portable fashion that results in performant code in three different architectures (CPU, GPU and Xeon Phi).  We also study its relevance to sparse linear algebra, partial differential equations, fine-grained hardware parallelism, and coarse-grained message-passing parallelism.


\section{Ensemble Propagation}
\label{sec:ensemble}
\begin{figure}
	\centering
	\subfigure[\label{fig:kron_sys}]{
		\includegraphics[width=0.45\textwidth]{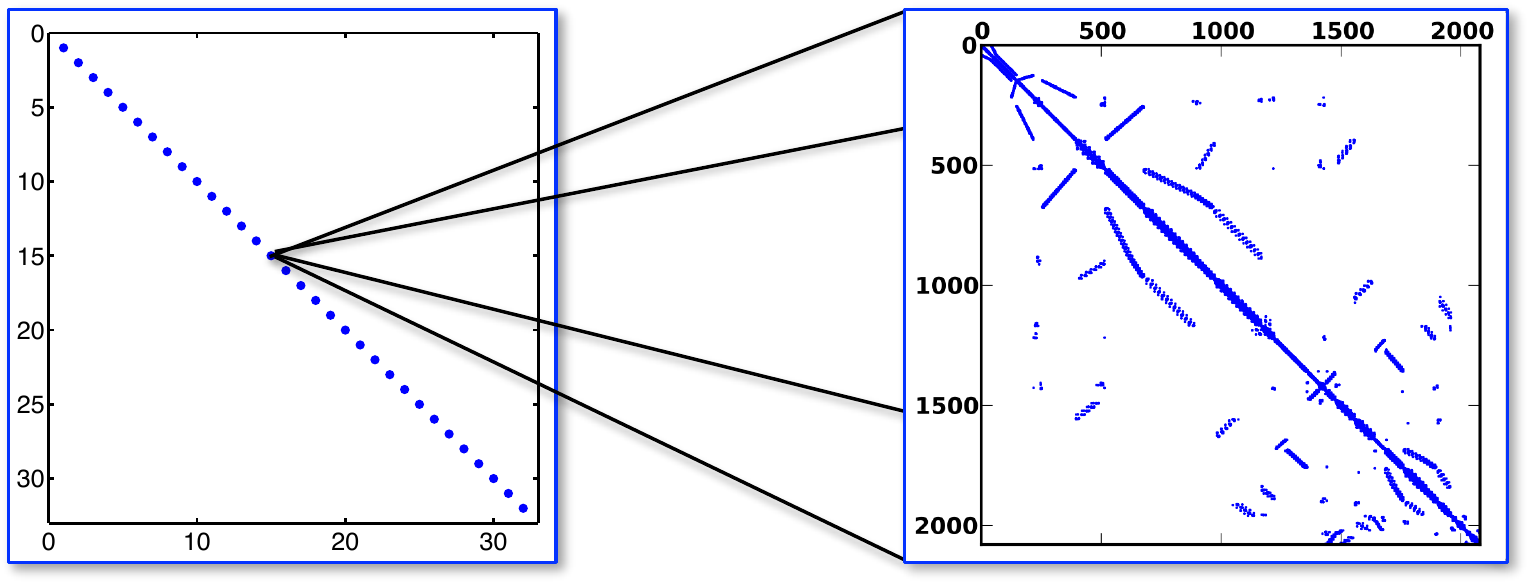}
	}
	\quad
 	\subfigure[\label{fig:kron_sys_commuted}]{
		\includegraphics[width=0.45\textwidth]{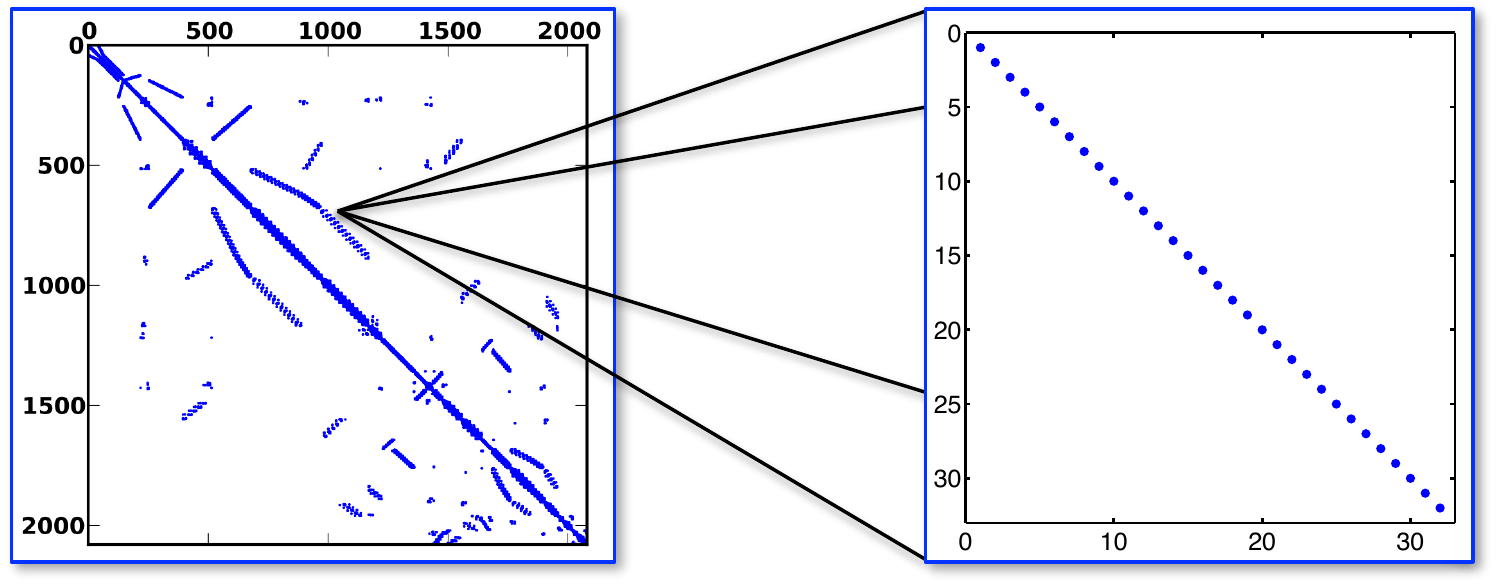}
	}
 	\caption{\subref{fig:kron_sys} Block diagonal structure of Kronecker product system~\eqref{eq:kron_sys}.  The number of blocks is determined by the ensemble size $s$, and each block has the sparsity structure for $\partial f/\partial u$.  \subref{fig:kron_sys_commuted}  Block structure of Kronecker product system~\eqref{eq:kron_sys_commuted}.  The outer structure is determined by $\partial f/\partial u$ where each nonzero is replaced by an $s\times s$ diagonal matrix.}
 	\label{fig:ensemble_matrix}
\end{figure}
In this section we formalize ensemble propagation and illustrate its effect on a ubiquitous computational kernel in the simulation of partial differential equations (PDEs), the sparse matrix-vector product.  For simplicity and brevity, we only consider steady-state problems here as the extension to transient problems is straightforward.  Consider a steady-state, finite-dimensional, nonlinear system
\begin{equation}
	f(u,y) = 0, \; u\in\R^n, \; y\in\R^m, \; f:\R^n\times \R^m\rightarrow\R^n.
\end{equation}
For the simulation of PDEs, we assume the equations have been spatially discretized by some suitable method (e.g., finite element, finite volume, finite difference), in which case $u$ would represent the nodal vector of unknowns, and $f$ the discretized PDE residual equations.  Here $y$ is a set of problem inputs, and we are interested in sampling the solution $u$ for numerous values of $y$.  Given some number $s$, consider computing $u$ for $s$ values of $y$:  $y_1,\dots,y_s$ (we assume $s$ is small and in what follows will be a small multiple of the natural vector width of the computer architecture).  Formally this can be represented by the Kronecker product system:
\begin{equation} \label{eq:kron_sys}
\begin{gathered}
	F(U,Y) = \sum_{i=1}^s e_i \otimes f(u_i,y_i) = 0, \\
	U = \sum_{i=1}^s e_i \otimes u_i, \; Y = \sum_{i=1}^s e_i \otimes y_i, \; 
\end{gathered}
\end{equation}
where $e_i\in\R^s$ is the $i$th column of the $s\times s$ identity matrix.  In this system, the solution vector $U$ is a block vector where all $n$ unknowns for each sample are ordered consecutively.  Furthermore, the Jacobian matrix $\partial F/\partial U = \sum_{i=1}^s e_ie_i^T \otimes \partial f/\partial u_i$ has a block diagonal structure, an example of which is shown in \Figref{fig:kron_sys}.

The choice of ordering for the unknowns in $U$ is arbitrary, and in particular the unknowns can be ordered so that all sample values are stored consecutively for each spatial degree of freedom in $u$.  Formally, this amounts to commuting the terms in the Kronecker product system:
\begin{equation} \label{eq:kron_sys_commuted}
\begin{gathered}
	F_c(U_c,Y_c) = \sum_{i=1}^s f(u_i,y_i) \otimes e_i = 0, \\
	U_c = \sum_{i=1}^s u_i \otimes e_i, \;\; Y_c = \sum_{i=1}^s y_i \otimes e_i.
\end{gathered}
\end{equation}
The Jacobian matrix $\partial F_c/\partial U_c = \sum_{i=1}^s \partial f/\partial u_i \otimes e_ie_i^T$ also has a block structure where each scalar nonzero in the original matrix $\partial f/\partial u$ is replaced by an $s\times s$ diagonal matrix, an example of which is shown in \Figref{fig:kron_sys_commuted}.

\begin{figure}
\begin{lstlisting}
// Matrix stored in compressed row storage for an arbitrary floating-point type T
template <typename T>
struct CrsMatrix {
    int num_rows;    // number of rows
    int num_entries; // number of nonzeros
    int *row_map;    // starting index of each row
    int *col_entry;  // column index for each nonzero
    T   *values;     // matrix values of type T
};

// CRS matrix-vector product z = A*x for arbitrary floating-point type T
template <typename T>
void crs_mat_vec(const CrsMatrix<T>& A, const T *x, T *z) {
    for (int row=0; row<A.num_rows; ++row) {
        const int entry_begin = A.row_map[row];
        const int entry_end   = A.row_map[row+1];
        T sum = 0.0;
        for (int entry=entry_begin; entry<entry_end; ++entry) {
            const int col = A.col_entry[entry];
            sum += A.values[entry] * x[col];
        }
        z[row] = sum;
    }
}
\end{lstlisting}
\caption{CRS matrix data structure and corresponding matrix-vector product routine.}
\label{fig:crs}
\end{figure}
To understand the performance aspects of the commuted ensemble system~\eqref{eq:kron_sys_commuted}, we will describe its impact on a ubiquitous kernel in the simulation of PDEs, the sparse matrix-vector product.  In \Figref{fig:crs}, a data structure for storing a matrix $A$ in the compressed row storage (CRS) format is displayed, along with a routine for computing matrix-vector products $z = Ax$ in this format.  Both the data structure and the matrix-vector product routine are templated to allow computations with any floating point scalar type.

Matrices and vectors for the ensemble system~\eqref{eq:kron_sys} can be stored in exactly the same format as in
\Figref{fig:crs}.  Only modifications for the matrix-vector product routine are required to add an additional
outer loop over the samples within the ensemble and adjustment of the indexing into {\tt A.values}, {\tt x} and {\tt z} as shown in \Figref{fig:outer_ensemble_crs}.
\begin{figure}
\begin{lstlisting}
// Ensemble matrix-vector product for commuted layout
template <typename T, int s>
void ensemble_crs_mat_vec(const CrsMatrix<T>& A, const T *x, T *z) {
    for (int e=0; e < s; ++e) {
        for (int row=0; row<A.num_rows; ++row) {
            const int entry_begin = A.row_map[row];
            const int entry_end   = A.row_map[row+1];
            T sum = 0.0;
            for (int entry=entry_begin; entry<entry_end; ++entry) {
                const int col = A.col_entry[entry];
                sum += A.values[e*A.num_entries+entry] * x[e*A.num_rows+col];
            }
            z[e*A.num_rows+row] = sum;
        }
    }
}
\end{lstlisting}
\caption{Matrix vector product routine for ensemble matrix and vectors in original layout.}
\label{fig:outer_ensemble_crs}
\end{figure}
Here the ensemble size {\tt s} is controlled through a compile-time constant template parameter.  This approach reduces memory requirements slightly since the row-offsets ({\tt A.row\_map}) and column-index ({\tt A.col\_entry}) arrays are only stored once for all matrices in the ensemble.  However for large matrices it is unlikely these arrays will fit in a low-level cache, and therefore these values are re-read for each sample within the ensemble.  Thus one would expect little improvement in performance.  

For the commuted Kronecker product
system~\eqref{eq:kron_sys_commuted} however, the ensemble loop is inside the row and column loops as is shown in
\Figref{fig:ensemble_crs}.  There are several
things to note.  First, any sample-dependent intermediate variables in the computation, such as {\tt sum}, are replaced by length-$s$ arrays.
Second, sample independent data, such as the row-offsets  and column-index arrays discussed above, are
read once per ensemble reducing aggregate
memory bandwidth requirements.  Third, a common approach for exploiting SIMD/SIMT vector parallelism within the
matrix-vector product calculation is to block the inner entry loop by some multiple of the vector width and replace the
multiply-add with the corresponding parallel instruction.  This requires a sparse gather for {\tt x} based on the indices
within {\tt A.col\_entry}.  The effectiveness of this depends on the number of nonzeros per row in the matrix, how these nonzeros are distributed across the matrix columns, and the capabilities of the computer architecture.  For the commuted ensemble system however, $s$ contiguous entries of {\tt x} are accessed consecutively from a common column offset {\tt col*s}.  Thus the ensemble loop around the multiply-add can be parallelized by SIMD/SIMT where the scalar sparse gather is replaced by a packed/coalesced load.  Furthermore, the ensemble loop is the inner-most loop, with a compile-time known trip-count and no dependencies between loop iterations, and thus is simple for the compiler to auto-vectorize.  Thus the effectiveness of applying SIMD/SIMT parallelism to the ensemble matrix-vector product routine effectively becomes independent of the matrix structure and the architecture's capabilities (assuming the ensemble size is chosen appropriately).  Finally, while not shown explicitly in \Figrefs{fig:crs}-\ref{fig:ensemble_crs}, the traditional approach for exploiting distributed-memory parallelism (e.g., MPI) within the matrix-vector product calculation is to distribute the rows of $A$, $x$ and $z$ across processors, where each processor computes the rows of $z$ it owns.  Based on the sparsity structure of $A$, this requires communicating entries of $x$ between processors so that all needed entries of $x$ are available to each processor, commonly referred to as the halo exchange.  When propagating one sample at a time, a halo exchange must occur for each matrix-vector product call, for each sample.  However for the ensemble system, only one halo exchange is necessary per ensemble, reducing the latency cost of this communication by a factor of $s$.
\begin{figure}
\begin{lstlisting}
// Ensemble matrix-vector product for commuted layout
template <typename T, int s>
void ensemble_crs_mat_vec(const CrsMatrix<T>& A, const T *x, T *z) {
    for (int row=0; i<A.num_rows; ++row) {
        const int entry_begin = A.row_map[row];
        const int entry_end   = A.row_map[row+1];
        T sum[s];
        for (int e=0; e < s; ++e)
            sum[e] = 0.0;
        for (int entry=entry_begin; entry<entry_end; ++entry) {
            const int col = A.col_entry[entry];
            for (int e=0; e < s; ++e) {
                sum[e] += A.values[entry*s + e] * x[col*s + e];
            }
        }
        for (int e=0; e < s; ++e)
            z[row*m + e] = sum[e];
    }
}
\end{lstlisting}
\caption{Matrix vector product routine for ensemble matrix and vectors in commuted layout.}
\label{fig:ensemble_crs}
\end{figure}

In short, the commuted Kronecker product system results in reduced aggregate bandwidth, packed loads/stores
(instead of sparse gather/scatter), better vector parallelization and reduced latency costs.
For this approach to be effective, one must choose the ensemble size $s$ appropriately for each architecture.  For effective SIMD/SIMT parallelism, it should be made a multiple of the natural vector width.  Increasing $s$ generally improves reuse and use of SIMD/SIMT hardware, at the expense of increased aggregate memory usage, cache pressure, and interconnect bandwidth pressure.  The effects of ensemble size choice for several architectures will be studied in \Secref{sec:results}.

The literature extensively discusses blocking solver algorithms to
improve performance.  In particular, authors have explored approaches
for exploiting block matrix structure in sparse matrix-vector product
algorithms to reduce memory bandwidth and increase instruction
throughput on superscalar architectures~\cite{im2004sparsity}.
Similarly, matrix-vector product kernels have been developed for
solving linear systems with multiple right-hand sides that leverage
reuse of the matrix graph and values across right-hand sides as well
as row-wise column ordering to apply SIMD/SIMT parallelism across
right-hand-side columns~\cite{im2004sparsity}.  The approach described
here is distinct in that it groups together normally independent
linear systems (each with a distinct right-hand side), and forms a
block solver algorithm from the constituent matrices and right-hand
sides.  It does not rely on any block structure within each matrix,
and can be viewed as an extension of the multiple right-hand side
approach to also include multiple left-hand sides.  Furthermore, the
necessary row-wise layout occurs naturally through the commuted
Kronecker product ordering.


\section{Embedded Implementation}
\label{sec:software}
In this section we discuss an \emph{embedded} approach for implementing ensemble propagation in general scientific simulation codes, based on C++ templates and operator overloading.  From this point on, we are only concerned with the commuted ensemble system~\eqref{eq:kron_sys_commuted}, henceforth refered to as the ``ensemble system.''  We first describe a new ensemble scalar type and a systematic approach for incorporating it into computer codes.  We then describe its integration with the manycore performance portability library Kokkos, and our solution to challenges in maintaining proper memory access patterns and exploiting vector parallelism.  Finally, we explain how we incorporate the technique into Trilinos' parallel linear algebra library called Tpetra~\cite{Baker:2012:SciProg,TpetraURL} and the iterative solvers and multigrid preconditioners that are built on Tpetra.

\subsection{Ensemble Scalar Type}
\label{sec:ensemble_type}
Converting the matrix-vector product routine to propagate ensembles of
values requires two fundamental steps: replace each sample-dependent
variable with a length-$s$ array, and replace each sample-dependent
arithmetic operation with a length-$s$ loop.  Implementing such a
change by hand is possible, but tedious and error prone.  Furthermore,
this is only one of many computational kernels and numerical
algorithms that preconditioned iterative linear solvers need.  Others
have explored using source-to-source transformation to automate this
conversion~\cite{Giering:2012}.  In this work we describe an approach
based on compile-time polymorphism, using C++ templates and operator
overloading.  The idea is to write application code and solvers that
are parametrized (``templated'') on the data type, and then replace
the customary floating-point data type (e.g., \texttt{double}) with
one that propagates ensembles.

\begin{figure}
\begin{lstlisting}
// Ensemble scalar type
template <typename T, int s>
class Ensemble {
    T val[s];
public:
    Ensemble(const T& v) {
        for (int e=0; e<s; ++e) val[m] = v;
    }
    Ensemble& operator=(const Ensemble& a) {
        for (int e=0; e<s; ++e) val[m] = a.val[m];
        return *this;
    }
    Ensemble& operator+=(const Ensemble& a) {
        for (int e=0; e<s; ++e) val[m] += a.val[m];
        return *this;
    }
    // ...
};

template <typename T, int s>
Ensemble<T,s>
operator*(const Ensemble<T,s>& a, const Ensemble<T,s>& b) {
    Ensemble<T,s> c;
    for (int e=0; e<s; ++e)
        c.val[e] = a.val[e]*b.val[e];
    return c;
}
\end{lstlisting}
\caption{Simplified ensemble scalar type definition.  Only the portions relevant to the CRS matrix-vector multiply routine are included, and complications such as expression templates are excluded.}
\label{fig:ensemble_class}
\end{figure}

We begin by defining the {\tt Ensemble} C++ class that stores an
ensemble of values.  \Figref{fig:ensemble_class} displays a portion of
it.  The \texttt{Ensemble} class, along with supporting utilities, is
provided by the Stokhos package\footnote{For expository purposes, we
  have simplified the {\tt Ensemble} class described here.  Stokhos'
  actual ensemble class differs in both interface and
  implementation.}~\cite{StokhosURL} within
Trilinos~\cite{TrilinosTOMS,TrilinosSP}.  It implements all of the
necessary operations for making objects of this type act like standard
floating-point data types, and for making them work with Trilinos'
data structures and solvers.  These operations fall into four
categories:
\begin{enumerate}
\item Arithmetic and basic mathematical functions
\item Input and output (C++ stream operations \texttt{<<} and \texttt{>>})
\item What Kokkos needs for reductions and atomic updates
\item What Tpetra needs to tell MPI how to communicate ensemble values
\end{enumerate}
The first category of operations make \texttt{Ensemble} objects act
like a C++ built-in floating-point type.  They include:
\begin{itemize}
\item Copy constructors and assignment operators from ensemble values
  as well as scalar values (that is, for objects of type
  \texttt{Ensemble<T,s>}, values of any type \texttt{U}
  convertible to \texttt{T})
\item Arithmetic operations ($+$, $-$, $\times$, $/$) and
  arithmetic-assignment operators ($+\!\!=$, $-\!\!=$, $\times\!\!=$,
  $/\!\!=$) from ensemble and scalar values
\item Comparison operations $>$, $<$, $==$, $<=$, and $>=$ between
  ensembles and scalar values, which base the result on the first
  entry in the ensemble
\item Overloads of basic mathematical functions declared in {\tt
    cmath}, such as \texttt{sin()}, \texttt{cos()},
  \texttt{exp()}, as well as other common operations such as
  \texttt{min()} and \texttt{max()}
\end{itemize}
The third category of operations lets Kokkos produce ensemble values
as the result of a parallel reduction or scan, and ensures that if
threads update the same ensemble value concurrently, those updates are
correct (assuming that order does not matter).  Kokkos needs overloads
of the above operations for ensemble values declared
\texttt{volatile}, and an implementation of Kokkos' atomic updates
for ensemble values.  \Secref{sec:solvers} will discuss the fourth
category of operations.

\texttt{Ensemble<T,s>} meets the C++ requirements for a \emph{plain
  old data} type.  The template parameter \texttt{s} fixes its size at
compile time; it does no dynamic memory allocation inside.
Furthermore, it has default implementations of the default
constructor, copy constructor, assignment operator, and destructor.
This implies that arrays of ensemble values can be allocated with no
more initialization cost than built-in scalars.  Ensemble objects can
be easily serialized for parallel communication and input / output,
since all such objects have the same size on all parallel processes.

Our ensemble scalar type employs expression
templates~\cite{VelhuizenET} to avoid creation of temporaries and fuse
loops within expressions, thus reducing overhead.  Since the ensemble
loop is always the lowest-level loop, it has a fixed trip count and no
iteration dependencies.  This means that the compiler can easily
auto-vectorize each arithmetic operation, including insertion of
packed load/store instructions.  However, difficulties do arise when
mapping GPU threads to ensemble components for GPU SIMT
parallelization.  We discuss our GPU optimizations in 
\Secref{sec:kokkos}.

Since the original CRS matrix-vector multiply routine in \Figref{fig:crs} is already templated on the scalar type, instantiating this code on the ensemble scalar type, {\tt crs\_mat\_vec< Ensemble<T,s> > >()} results precisely in the ensemble matrix-vector multiply routine in \Figref{fig:ensemble_crs}.  

\subsection{Incoporating the Ensemble Type in Complex Codes}
\label{sec:ensemble_complex_codes}
Replacing the floating-point scalar type with the ensemble types accomplishes both steps of replacing sample-dependent
variables with ensemble arrays and arithmetic operations with ensemble loops.  We generally advocate the use of C++ templates
to facilitate this type change, whereby the floating-point type is replaced by a general template parameter.  This allows the
original code to be recovered by instantiating the template code on the original floating-point type, and the ensemble code
through the ensemble scalar type.  Furthermore, other scalar types can be used as well, such as automatic differentiation
types for computing derivatives.  We refer to the process of using C++ templates and a variety of scalar types to implement
different forms of analysis as \emph{template-based generic programming}~\cite{Pawlowski:2012kc,Pawlowski:2012js}.
This has been shown to be effective for supporting analyses such as ensemble propagation in complex simulation codes.

The most challenging part of incorporating the ensemble scalar type in complex code bases is its conversion to code templated on the scalar type.  Developers must analyze the code to determine which values depend (directly or indirectly) on the input data that will be sampled, and therefore should be converted to ensembles by replacing the types of those values with a template parameter.  While this is admittedly tedious, it is generally straightforward to accomplish.  Furthermore, the compiler helps in this process since the ensemble scalar type does not allow direct conversions of ensemble values to scalar values.  This prevents accidentally breaking the chain of dependencies from input data to simulation outputs.   It is possible to implement this conversion through a helper function, which takes the first entry in the ensemble to initialize the resulting scalar.  Therefore one can incrementally convert a code to use ensembles by manually converting ensembles to scalars whenever ensemble code calls code that has not yet been converted.  Note that scalars can be automatically converted to ensemble values by the compiler, which implies it is possible to incorrectly replace code that does not depend on the input data with ensembles.  This can only be discovered through programmer analysis and optimization of the code.

Once all necessary scalar values have been replaced by ensembles in the simulation code, the ensemble propagation occurs
automatically by ``evaluating'' the resulting ensemble code.  This requires adding suitable initialization and finalization code to initialize ensemble values for input data and extracting ensemble values for simulation results (using various ensemble constructors and coefficient access routines).  Sample independent data are not replaced by ensembles, and
therefore reuse happens naturally through the normal compiler optimization process and overloaded operators that take a
mixture of scalars and ensembles as arguments.  


\subsection{Kokkos Performance Portability}
\label{sec:kokkos}

Kokkos~\cite{Kokkos:2012:SciProg,Kokkos:2014:JPDC} is a programming model and C++ library that enables applications and domain libraries to implement thread scalable algorithms that are performance portable across diverse manycore architectures such as multicore CPU, Intel Xeon Phi, and NVIDIA GPU. 
Kokkos' design strategy is to define algorithms with parallel patterns (for-each, reduction, scan, task-dag) and their code bodies invoked within these patterns, and with multidimensional arrays of their ``scalar'' data types.
Performance portability is realized through the integrated mapping of patterns, code bodies, multidimensional arrays, and datum onto the underlying manycore architecture.

This mapping has three components.
First, code is mapped onto the target architecture's best performing threading mechanism; \textit{e.g.}, pthreads or OpenMP on CPUs and CUDA on NVIDIA GPUs.
Second, parallel execution is mapped with architecture-appropriate scheduling; \textit{e.g.}, each CPU thread is given a contiguous range of the parallel iteration space while each GPU thread is given a thread-block-strided range of the parallel iteration space.
Third, multidimensional arrays are given an architecture-appropriate layout; \textit{e.g.}, on CPUs arrays have a row-major or ``array of structs'' layout and on GPUs arrays have a column-major or ``struct of arrays'' layout.
While this polymorphic multidimensional array abstraction has conceptual similarities to Boost.MultiArray~\cite{website:Boost:MultiArray}, Kokkos' abstractions for explicit dimensions and layout specializations provides greater opportunities for performance optimizations.

When scalar types are replaced with ensemble types in a Kokkos multidimensional array, the layout is specialized so that operations on ensemble types may exploit the lowest level of hierarchical parallelism.
Hierarchical thread parallelism can be viewed as ``vector'' parallelism nested within thread parallelism.
The mechanism to which Kokkos maps ``vector'' parallelism is architecture dependent.
On CPUs this level is mapped to vector instructions, typically through the compiler's optimization algorithms.
On GPUs this level is mapped to threads within a GPU warp, and then Kokkos' thread abstraction is remapped to the entire warp.
Thus on the GPU architecture ensemble operations are, transparent to the user code, performed in parallel by a warp of threads.

The specialized layout integrates the ensemble dimension into the multidimensional array such that the ensemble's values remain contiguous in memory on any architecture.
This contiguity is necessary to obtain the best ``vector'' level parallel performance for computations on irregular data structures; such as sparse matrices and unstructured finite elements.
These data structures typically impede performance by requiring non-contiguous scalar values to be gathered into contiguous memory (vector registers), processed with vector instructions, and then results scattered back to non-contiguous scalar values.
On CPU architectures these gather/scatter operations \textit{might} be automatically generated by a compiler that recognizes the indirection patterns of irregular data structures.
When the scalar type is an ensemble each indirect access that previously referenced a single scalar value instead references the ensemble types' contiguous set of values (recall \Figref{fig:ensemble_class}).
As such, gather/scatter operations are no longer needed for vector instructions, and compilers can more easily generate vectorized operations.

On NVIDIA GPU architectures, warp-level gather/scatter operations are generated by hardware, removing the need for sophisticated indirection-pattern recognition by the compiler.
However, memory accesses are still non-coalesced gather/scatter operations thus reducing performance.
When the scalar type is an ensemble, indirect access patterns lead to coalesced reads and writes of contiguous ensemble values, resulting in improved performance.

In summary, replacing scalar types with ensemble types in computations on irregular data structures enables improved utilization of hierarchical parallel hardware such as multicore CPUs with vector instructions and GPUs.
To realize this improvement
(1) Kokkos multidimensional array layouts are specialized to insure ensemble values are contiguous in memory and
(2) ensemble operations are mapped to the ``vector'' level of Kokkos' hierarchical thread-vector parallelism.
On CPU architectures, this mapping happens automatically through the normal compiler vectorization process when applied to the ensemble loops.  For GPU architectures however, this mapping occurs by creating a strided subview within the ensemble dimension of the multidimensional array for each GPU thread within a warp.
These two mappings insure that CPU vector instructions or GPU warp operations are performed on contiguous memory.


\subsection{Linear Algebra and Iterative Solvers}
\label{sec:solvers}

The ensemble scalar type and Kokkos library described above have both
been incorporated into the Tpetra linear algebra
package~\cite{Baker:2012:SciProg,TpetraURL} within Trilinos.  Tpetra
implements parallel linear algebra data structures, computational
kernels, data distributions, and communication patterns.  ``Parallel''
includes both MPI (the Message Passing Interface) for
distributed-memory parallelism, and Kokkos for shared-memory
parallelism within an MPI process.  Supported data structures include
vectors, ``multivectors'' that represent groups of vectors with the
same parallel distribution, sparse graphs, sparse matrices, and
``block'' sparse matrices (where each block is a small dense matrix).
Tpetra's computational kernels include vector arithmetic, sparse
matrix-vector products, sparse triangular solve, and sparse
matrix-matrix multiply.  It lets users represent arbitrary
distributions of data over MPI processes, and communicate data between
those distributions.

Tpetra is templated on the ``scalar'' type, the type of each entry in
the matrix or vector.  In theory, this lets Tpetra work with any type
that ``looks like'' one of the C++ built-in floating-point types.
Tpetra uses this flexibility to provide built-in support for both
single- and double-precision real and complex floating-point values,
as well as 128-bit real floating-point arithmetic if the compiler
supports it.  In practice, in order to support an arbitrary scalar
type, Tpetra needs it to implement all of the operations described in
\Secref{sec:ensemble_type}.  In particular, Tpetra needs to tell MPI
how to communicate scalars.  This means that Tpetra either needs to
know the \texttt{MPI\_Datatype} corresponding to the scalar type, or
how to pack and unpack scalars into byte arrays.  The scalar type
tells Tpetra this by implementing a C++ traits class specialization.
Tpetra in turn provides a type-generic MPI interface, both for itself
and for users.

The Belos package~\cite{Bavier:2012:SciProg} in Trilinos builds upon Tpetra data structures to provide parallel iterative linear solvers such as CG and GMRES.  It is also templated on the scalar type, allowing ensembles to be propagated through these linear solver algorithms.  Iterative solver algorithms do not directly access matrix and vector entries.  They only need to know the results of inner product and norm calculations, and only deal with matrices and vectors as abstractions.  Furthermore, viewing the ensemble system as the Kronecker product system~\eqref{eq:kron_sys_commuted}, inner products and norms of ensemble vectors should be scalars and not ensembles.  That is, given two ensemble vectors $U_c = \sum_{i=1}^s u_i \otimes e_i$ and $V_c = \sum_{i=1}^s v_i \otimes e_i$,
\begin{equation}
	U_c^T V_c = \sum_{ij=1}^s u_i^T v_j \otimes e_i^Te_j = \sum_{i=1}^s u_i^T v_i.
\end{equation}
Tpetra assists Belos by defining a traits class that defines the proper data type for the result of inner product and norm calculations, and exposing these types to solvers and application code as public typedefs.  First
\begin{equation}
	\sum_{i=1}^s u_i^T v_i \otimes e_i
\end{equation}
is computed, which naturally arises when propagating the ensemble scalar type through the inner product code.   The final inner product value is then computed by adding together each ensemble component.  Thus, Belos does not directly generate or access ensemble values; they only appear internally within the matrix and vector data structures.  Belos only needs one implementation for Tpetra objects of all scalar types, with no significant abstraction overhead.

\subsection{Multigrid Preconditioners}
\label{sec:preconditioners}

\begin{algorithm}[t]
\centering
\begin{algorithmic}[0]
  \State{$A^0 = A$}
  \Function{VCycle}{$A^k$, $b$, $x$, $k$}
    \State{// Solve $A^k$ x = b (k is current grid level)}
    \State $ x = S^{k}_{1} (A^k, b, x)$
      \If{$(k \ne {\bf N-1})$}
        \State{$r^{k+1} = R^k (b - A^k x )$}
        \State{$A^{k+1} = R^k A^k P^k$}
        \State{$z = 0$}
        \State{}\Call{VCycle}{$A^{k+1}$, $r^{k+1}$, $z$, $k+1$}
        \State{$ x = x + P^{k} z$}
        \State{$ x = S^{k}_{2} (A^k, b, x )$}
      \Else
        \State{$x = S^{c}(A^{N-1}, x,  b)$}
      \EndIf
  \EndFunction
\end{algorithmic}
\caption{V-cycle multigrid with $N$ levels to approximate solution $Ax=b$.}
\label{vcycle}
\end{algorithm}
Multigrid is a provably optimal linear solution method in work per digit of accuracy for
systems arising from elliptic PDEs.
Multigrid works by accelerating the solution of a linear system of interest, $A^0x^0=b^0$, through a
sequence or {\em hierarchy} of increasingly smaller linear systems $A^ix^i=b^i, i>0$. The purpose of
each system or {\em level} is to reduce particular ranges of errors in the $i=0$ problem.  Any errors that are not
quickly damped by a particular system should be handled by a coarser problem.

The main components of a multigrid solver are {\em smoothers}, which are solvers that operate only on
particular levels, and transfer operators to migrate data between levels.
The transfer from level $i+1$ to $i$ is called a {\em prolongator} and denoted $P^i$.
The transfer  from level $i$ to $i+1$ is the {\em restrictor} and denoted $R^i$.
A typical schedule for visiting the levels, called a {\em V-cycle}, is given in Algorithm~\ref{vcycle}.  In practice, this
algorithm is divided into two phases:  the setup phase where all of the matrix data used at each level is generated (e.g.,
$R^k$, $P^k$, $A^k$, $S^k_1$, $S^k_2$ and $S^c$), and the solve phase where given $b$ and the data from the setup phase, $x$ is computed.

Multigrid algorithms such as this have been implemented in the MueLu package within Trilinos~\cite{muelu_user}. MueLu's performance with traditional scalar types on very large
 core counts has been studied before~\cite{LinLPP14,LinPPL}. This library builds upon the
templated Tpetra data structures and algorithms described above, with all of the functionality used in setup (such as the
matrix-matrix multiply $R^k A^k P^k$) and application of the V-cycle in Algorithm~\ref{vcycle} (such as the smoothers
$S^{k}_1$ and $S^{k}_2$) templated on the scalar type.  Furthermore, the V-cycle algorithm in MueLu is encapsulated within an operator $z = M x$ allowing it to serve as a preconditioner for the Krylov methods in Belos.  In this work, the restriction and prolongation operators $R^k$ and $P^k$ are generated from the graph of $A^k$ at each level, without any thresholding or dropping.  This implies that given matrices $A_1,\dots,A_s$ corresponding to $s$ samples within an ensemble, and corresponding multigrid preconditioners $M_1,\dots,M_s$ generated for each matrix {\em individually}, then the corresponding ensemble preconditioner $M_c$ is equivalent to
\begin{equation}
	M_c = \sum_{i=0}^s M_i \otimes e_ie_i^T.
\end{equation}
In the experiments described below, we use order-2 Chebyshev polynomial smoothers ($S_1^k$, $S_2^k$) and continue to generate levels in the multigrid hierarchy until the number of matrix rows falls below 500.  For the coarse-grid solve $S^c$, a simple sparse-direct solver that is built into Trilinos, called Basker, is used.  Note that in the current version of the code, the preconditioner setup and coarse grid solver do not leverage Kokkos directly and thus are thread-serial (but are MPI-parallel). Furthermore, for the GPU architecture, the coarse-grid solve is executed on the host using recent Unified Virtual Memory (UVM) features to automatically transfer data between the host and GPU.  All other aspects of the V-cycle solve phase are fully thread-parallel using Kokkos.

A major concern in parallel multigrid is the relative cost of communication to
computation for levels $i>0$.  For a good-performing multigrid method, matrix
$A^{i+1}$ can typically have 10--30 times fewer rows than $A^i$, with only a
modest increase in the number of nonzeros per row.  Practically, this means
that the ratio of communication to computation can increase by an order of
magnitude per level.
In MueLu, this issue is addressed by moving data
to a subset of processes for coarser levels.  Once, the number of processes
for a coarser level is determined, a new binning of the matrix rows
(weighted by the number of nonzeros per row) is found using the
multi-jagged algorithm~\cite{mj_tpds} from the Zoltan2 library~\cite{zoltan2}.
Each bin is assigned to a process so that data movement is restricted in the
coarser level to fewer processes.  
This improves scalability of the preconditioner, particularly with large numbers of processes.  Propagating ensembles through the preconditioner further reduces communication costs by amortizing communication latency across the ensemble.    

\subsection{Build times and library sizes}
\label{sec:ETI}

Some developers worry that extensive use of C++ templates may increase
compilation times and library sizes.  The issue is that plugging each
scalar type into templated linear algebra and solver packages results
in an entirely new version of the solver to build.  The compiler sees
a matrix-vector multiply with \texttt{double} as distinct code from a
matrix-vector multiply of \texttt{Ensemble<T,s>}, for example.  More
versions of code means longer compile times and larger libraries.
This is not particular to C++ templates; the same would occur when
implementing ensemble computations automatically using
source-to-source translation with Fortran or C, with manual
translation, or with some other language's flavor of compile-time
polymorphism.

A second issue is particular to C++.  Most C++ compilers by default
must re-build templated code from scratch in each source file that
uses it.  Furthermore, deeply nested ``stacks'' of solver code do not
actually get compiled until an application source file uses them with
a specific scalar type.  For example, MueLu is templated and depends
on many Trilinos packages that are also templated, so using MueLu
means that the application must build code from many different
Trilinos packages.  In practice, this shows up as long application
build times, or even running out of memory during compilation.

Trilinos fixes this with its option to use what it calls
\emph{explicit template instantiation} (ETI).  This ``pre-builds''
heavyweight templated code so applications do not have to build it
from scratch each time.  ETI corresponds to the second option in
Section 7.5 of the GCC Manual \cite{gcc52manual}, where Trilinos
manually instantiates some of its templated classes.  
ETI means that building
Trilinos might take longer, but building the application takes less
time.  Trilinos breaks up many of its instantiations of templated
classes and functions into separate source files for different
template parameters, which keeps down build times and memory
requirements for Trilinos itself.  


\subsection{Ensemble Divergence}
\label{sec:divergence}
The most significant algorithmic issue arising from the embedded ensemble propagation approach described above is \emph{ensemble divergence}.  Depending on the values of two given samples, the code paths taken during evaluation of the simulation code at those two samples may be different.  These code paths must some how be joined together when those samples are combined into a single ensemble.  We now describe different approaches for accomplishing this depending on how and where the divergence occurs within the simulation code.
\begin{figure}
\begin{lstlisting}
// ...

Scalar x = ...

Scalar y;
if (x > 0) {
   Scalar z = x*x;
   y = x + z;
else
  y = x;
  
// ...
\end{lstlisting}
\caption{Example of ensemble divergence due to conditional evaluation resulting in multiple code branches.  Here {\tt Scalar} is a general template parameter that could be {\tt double} for single-point evaluation or {\tt Ensemble} for ensemble evaluation.}
\label{fig:conditional}
\end{figure}
\begin{figure}[h]
\begin{lstlisting}

// Base template definition of EnsembleTrait that is empty.
// It must be specialized for each scalar type T
template <typename T> struct EnsembleTrait {};

// Specialization of EnsembleTrait for T = double
template <> struct EnsembleTrait<double> {
  typedef double value_type;
  static const int ensemble_size = 1;
  static const double& coeff(const double& x, const int i) { return x; }
  static       double& coeff(      double& x, const int i) { return x; }
};

// Specialization of EnsembleTrait for T = Ensemble
template <typename T, int s> struct EnsembleTrait< Ensemble<T,s> > {
  typedef T value_type;
  static const int ensemble_size = s;
  static const T& coeff(const Ensemble<T,s>& x, const int i) { return x.val[i]; }
  static       T& coeff(      Ensemble<T,s>& x, const int i) { return x.val[i]; }
};

// ...

typedef EnsembleTrait<Scalar> ET;
typedef typename ET::value_type ScalarValue;
const int s = ET::ensemble_size;

Scalar x = ...

Scalar y;
for (int i=0; i<s; ++i) {
    const ScalarValue& xi = ET::coeff(x,i);
          ScalarValue& yi = ET::coeff(y,i);
    if (xi > 0) {
       ScalarValue z = xi*xi;
       yi = xi + z;
    }
    else
      yi = xi;
      
// ...
}
\end{lstlisting}
\caption{Handling ensemble divergence through the {\tt EnsembleTrait} type trait which enables loops over ensemble components in a type-generic fashion.  }
\label{fig:ensemble_conditional}
\end{figure}

When divergence occurs at low levels within the simulation code, for example during element residual or Jacobian evaluation of the PDE, a simple approach for handling it is to add a loop over ensemble components that evaluates each sample individually.  An example of this is demonstrated in \Figrefs{fig:conditional} and~\ref{fig:ensemble_conditional}.  In \Figref{fig:conditional}, a code branch is chosen based on the value of {\tt x}, whose type is determined by the template parameter {\tt Scalar}.  When {\tt Scalar} is a basic floating-point type such as {\tt double} for a single-point evaluation, everything is fine.  However when {Scalar} is {\tt Ensemble<T,s>}, only one of the branches can be chosen even when the components of {\tt x} would choose different branches when evaluated separately.  This is remedied in \Figref{fig:ensemble_conditional} by adding a type-generic loop around the conditional, evaluating the loop body separately for each sample within the ensemble.  This recovers the single-point behavior.  This is accomplished through the type trait {\tt EnsembleTrait<T>} displayed in \Figref{fig:ensemble_conditional} which has a trivial implementation for built-in types such as {\tt double} allowing the code to be instantiated for both {\tt double} and {\tt Ensemble}.  Clearly, the use of an ensemble loop such as this eliminates the architectural benefits of ensemble propagation through the body of the loop, and therefore should only be applied to small portions of the code where the bodies of the conditional branches are small.

Divergence may also occur at high levels within the simulation code.
Examples include iterative linear and nonlinear solver algorithms that
require a different number of solver iterations for each sample, and
adaptive time stepping and meshing schemes that dynamically adjust the
temporal and spatial discretizations based on error estimates computed
for previous time steps or solutions.  While adding an ensemble loop
around these calculations is certainly feasible, it defeats the
original intent of incorporating ensemble propagation.

Alternately, recall that the use of an ensemble scalar type is merely
a vehicle for implementing the Kronecker product
system~\eqref{eq:kron_sys_commuted}.  As we discussed in
\Secref{sec:solvers}, the norm and inner product calculations that
drive convergence decisions for iterative solver algorithms as well as
adaptivity decisions for time-stepping and meshing do not produce
ensemble values.  Instead, they produce traditional floating-point
values, which are the results of norms and inner products over entire
ensemble vectors.  This effectively \emph{couples} all of the ensemble
systems together, resulting in a single convergence or adaptivity
decision for the entire ensemble system.  Thus divergence is handled
in these cases through proper definition of the types used for
magnitudes and inner products, with associated traits classes for
computing these quantities in a type-generic fashion.  The resulting
coupled linear solver algorithm is analogous to block Krylov subspace
methods~\cite{oleary1980block}.  First, it couples the linear systems
together \emph{algorithmically}, not just computationally.  Second,
our approach opens up opportunities for increasing spatial locality
and reuse in computational kernels, just as block Krylov methods do.
(See, for example, \cite{baker2006efficient}).  Just as with block
Krylov methods, however, coupling the component systems comprising the
ensemble system means that the solves are no longer algorithmically
equivalent to uncoupled solves.  This means that the choice of the
ensemble size $s$, as well as which samples are grouped together
within each ensemble, will affect the performance of the resulting
simulation algorithms.
Therefore, the solution to managing high-level solver divergence across ensemble values is group samples together in ensembles that minimize this divergence.

For example, the convergence of iterative linear solvers (or its number of
iterations) for each sample depends on several factors.  Among these,
the most important are the condition number of the matrix associated
with $\partial f/\partial u$ and the spatial variation and magnitude
of the sample-dependent parameters. When the ensemble system $F_c$ is
formed, the solver's convergence is always poorer than the solver
applied to each sample individually. This happens because the spectrum
of the ensemble matrix is the union of the matrix spectra of the
samples that comprise it; this likely increases the condition number
and hence the number of iterations. For this reason, it is convenient
to have a grouping strategy that gathers samples requiring a similar
number of iterations in the same ensemble. Since this information is
not known {\it a priori}, quantities such as those mentioned above can
be used to predict which samples feature a similar number of
iterations.  Preliminary studies show that the variation of the
sample-dependent parameters over the computational domain may induce a
grouping very similar to the one based on the number of iterations.
Algorithmic ideas along these lines will be explored in a future
paper.  In the computational studies below however, we assume the
matrix spectra for all samples are similar so that the number of
iterations of the ensemble system is constant with regards to
grouping.


\section{Computational Results}
\label{sec:results}
We now investigate the performance improvements possible for the ensemble propagation approach described above.  We consider the simple nonlinear advection-diffusion system
\begin{equation}\label{eq:diffusion_sys}
	-\nabla \cdot (\kappa(x,y)\nabla u) + \alpha v\cdot\nabla u + \beta u^2 = 0, \; x\in D = [0,1]^3.
\end{equation}
The diffusion coefficient $\kappa$ is treated as uncertain and is represented by a truncated \KL expansion~\cite{Ghanem_Spanos_91}:
\begin{equation}
	\kappa(x,y) = \kappa_0 + \sigma \sum_{i=1}^m \kappa_i(x)y_i, \; x\in D, \; y\in\Gamma = [-1,1]^m,
\end{equation}
where each $y_i$ is a uniform random variable over $[-1,1]$, $\kappa_0$ is the mean of the diffusion coefficient, $\sigma$ its standard deviation, the $\kappa_i$ are eigenfunctions of the exponential covariance function
\begin{equation}
	C(x,x') = \sigma^2 e^{-\frac{\|x-x'\|_1}{L}}, \; x,x'\in D,
\end{equation}
and $L$ is the correlation length (in this case, the $\kappa_i$ are tensor products of sines and cosines with frequencies that grow with $i$~\cite{Ghanem_Spanos_91}).  We discretize \Eqref{eq:diffusion_sys} using linear finite elements, resulting in a discrete nonlinear system $f(u,y) = 0$.  For each sample of $y$, we solve this system iteratively via Newton's method, which at each iteration requires computing the solution to the linear system
\begin{equation} \label{eq:fenl_lin_sys}
	A\Delta u = -f, \;\; A = \frac{\partial f}{\partial u}.
\end{equation}
We assemble and solve \Eqref{eq:fenl_lin_sys} via a hybrid distributed-shared memory parallel approach where given the number
of MPI ranks $p$, the spatial domain $D$ is divided into $p$ disjoint subdomains resulting in a partitioning of
\Eqref{eq:fenl_lin_sys} into $p$ disjoint sets of rows.  The processors associated with each MPI rank then assemble and solve
the equations belonging to them using a shared-memory parallel approach implemented by Kokkos described below.  
\begin{algorithm}
\centering
\renewcommand{\algorithmiccomment}[1]{// #1}
\begin{algorithmic}[0]
\Function{Residual\_Jacobian\_Assembly}{$A$, $f$, $u$, $y$}
\State \Comment{Import needed off-processor elements of $u$}
\State halo\_exchange($u$) 
\State
\State \Comment{Thread-parallel for loop over mesh cells}
\For{$e=0$ to $N_{mesh}$}
  \State
  \State \Comment{Sparse gather of element solution vector}
  \For{$i=0$ to $N_{node}$} 
    \State $I$ = NodeIndex($e$,$i$)
    \State $u_e(i)$ = $u$($I$)
  \EndFor
  \State
  \State \Comment{Evaluate element residual $f_e$ and Jacobian $A_e$}
  \For{$qp=0$ to $N_{qp}$}
  	\State\Comment{Evaluate basis functions, gradients, and}
	\State\Comment{transformation from reference cell at}
	\State\Comment{quadrature point $qp$}
	\State
	\State\Comment{Evaluate element solution $u_e$ and}
	\State\Comment{gradient $\nabla u_e$ at quadrature point}
	\State
	\State\Comment{Evaluate diffusion, advection, source terms}
	\State\Comment{and sum into element residual/Jacobian}
  \EndFor
  \State
  \State \Comment{Sparse scatter into global residual/Jacobian}
  \For{$i=0$ to $N_{node}$}
    \State $I$ = NodeIndex($e$,$i$)
    \State atomic\_add($f$($I$),  $f_e(i)$) 
    \For{$j=0$ to $N_{node}$}
      \State $J$ = ElemGraph($e$,$i$,$j$)
      \State atomic\_add($A$($I$,$J$),  $A_e(i,j)$)
    \EndFor
  \EndFor
\EndFor
\EndFunction
\end{algorithmic}
\caption{Pseudo-code description of PDE assembly algorithm.}
\label{alg:assembly}
\end{algorithm}

A pseudocode description of the assembly portion of the calculation is presented in \Algreff{alg:assembly}.  Given values for $u$ and $y$, the residual $f(u,y)$ and Jacobian $\partial f/\partial u = A$ are evaluated.  Here $N_{mesh}$ is the number of mesh cells for a given MPI rank, $N_{node} = 8$ is the number of finite element basis functions/vertices per cell, and $N_{qp} = 8$ is the number of quadrature points per cell.  First entries of $u$ owned by other MPI ranks that are needed within each MPI rank's assembly are imported in the halo exchange.  This communication pattern is precisely the same as needed in sparse matrix-vector products involving $A$.  Then each element (cell) in the mesh belonging to the given MPI rank is visited, where local contributions to the global residual and Jacobian are computed.  This loop is executed in parallel by Kokkos, where each cell is assigned a unique thread.  A mapping from mesh nodes to global degree-of-freedom indices (NodeIndex) is used to initialize the local element solution vector $u_e$.  The element residual and Jacobian are computed through a numerical integration which requires evaluating the finite element basis functions, their gradients, the corresponding solution vector $u_e$, the gradient of the solution vector $\nabla u_e$, and the PDE diffusion, advection, and source terms at each quadrature point (not shown for brevity).  The results of these calculations are accumulated into the local residual and Jacobian.  Finally the element residual and Jacobian contributions are accumulated into the global residual and Jacobian data structures using the NodeIndex mapping as well as a second mapping of element Jacobian columns to global matrix columns (ElemGraph).  Since multiple cells contribute to the same residual/Jacobian row, these contributions must use atomic instructions to prevent thread race conditions.  The C++ code implementing the assembly algorithm (\ref{alg:assembly}) is templated on the scalar type as described above, using templated Tpetra matrix and vector data structures for storing $f$ and $A$.  This allows reuse of the same templated code base for both scalar (single sample value) and ensemble systems.

In the experiments below, we choose $\alpha = \beta = 0$, in which case the resulting discrete equations are linear, symmetric, and positive-definite.  Thus only a single iteration of Newton's method is required, and the Newton system \Eqref{eq:fenl_lin_sys} is solved by CG preconditioned by multigrid as described in
\Secref{sec:preconditioners}.  We also choose $m = 5$, $\kappa_0 = 1$, $\sigma = 0.1$, and $L = 1$, in which case the diffusion coefficient varies smoothly with respect to both $x$ and $y$, the number of CG iterations is very nearly constant with respect to samples of $y$, and the number of CG iterations does not vary with $s$.

\begin{table*}[t]
  \centering
    \begin{tabulary}{\textwidth}{lLLLl}
    \toprule
    Name & Processor Description & Interconnect Description & Thread Parallelism & Compiler \\
    \midrule
    Sandy Bridge & Dual-socket 2.6 GHz, Intel Xeon E5-2670 CPU & Infiniband QDR    & OpenMP & Intel 14.0 \\
    Blue Gene/Q &16 core, IBM PowerPC A2 CPU & 5-D Torus & OpenMP & GNU 4.7.2 \\
    Cray XK7 & 16-core, 2.2 GHz, AMD Interlagos CPU & 3-D Torus & OpenMP & GNU 4.8.2 \\
    GPU & NVIDIA K20x GPU & Infiniband QDR & CUDA & NVCC 6.5 \\
    Accelerator & Intel Xeon Phi 7120P, 60 cores, 4 threads/core & N/A & OpenMP & Intel 15.0 \\
    \bottomrule
    \end{tabulary}
    \caption{Computational architectures studied.}
  \label{table:arch}
\end{table*}
Given an ensemble size $s$, we measure speed-up of the ensemble approach for assembling and solving \Eqref{eq:fenl_lin_sys} for $s$ values of $y$ simultaneously over the scalar approach of assembling and solving this system $s$ time sequentially, by measuring the timings of the two approaches implemented on the parallel architectures\footnote{Due to insufficient support for the C++98 standard by the native IBM compilers, the GNU compilers were used on the Blue Gene/Q architecture, disallowing the use of vector instructions.  At the time of this writing, performance issues  associated with the CUDA toolkit available for the Cray XK7 prohibited generating computational results with this code base on the GPU.  Therefore GPU results were obtained on a small cluster.  Also, only a single Intel Xeon Phi accelerator card is available to the authors.} displayed in \Tableref{table:arch}.  Here speed-up is defined as
\begin{equation}
	\mbox{Speed-Up} = \frac{s 
	\times \mbox{Time for single sample}}{\mbox{Time for ensemble}}.
\end{equation}

\begin{figure}[t]
	\centering
	\includegraphics[width=0.45\textwidth]{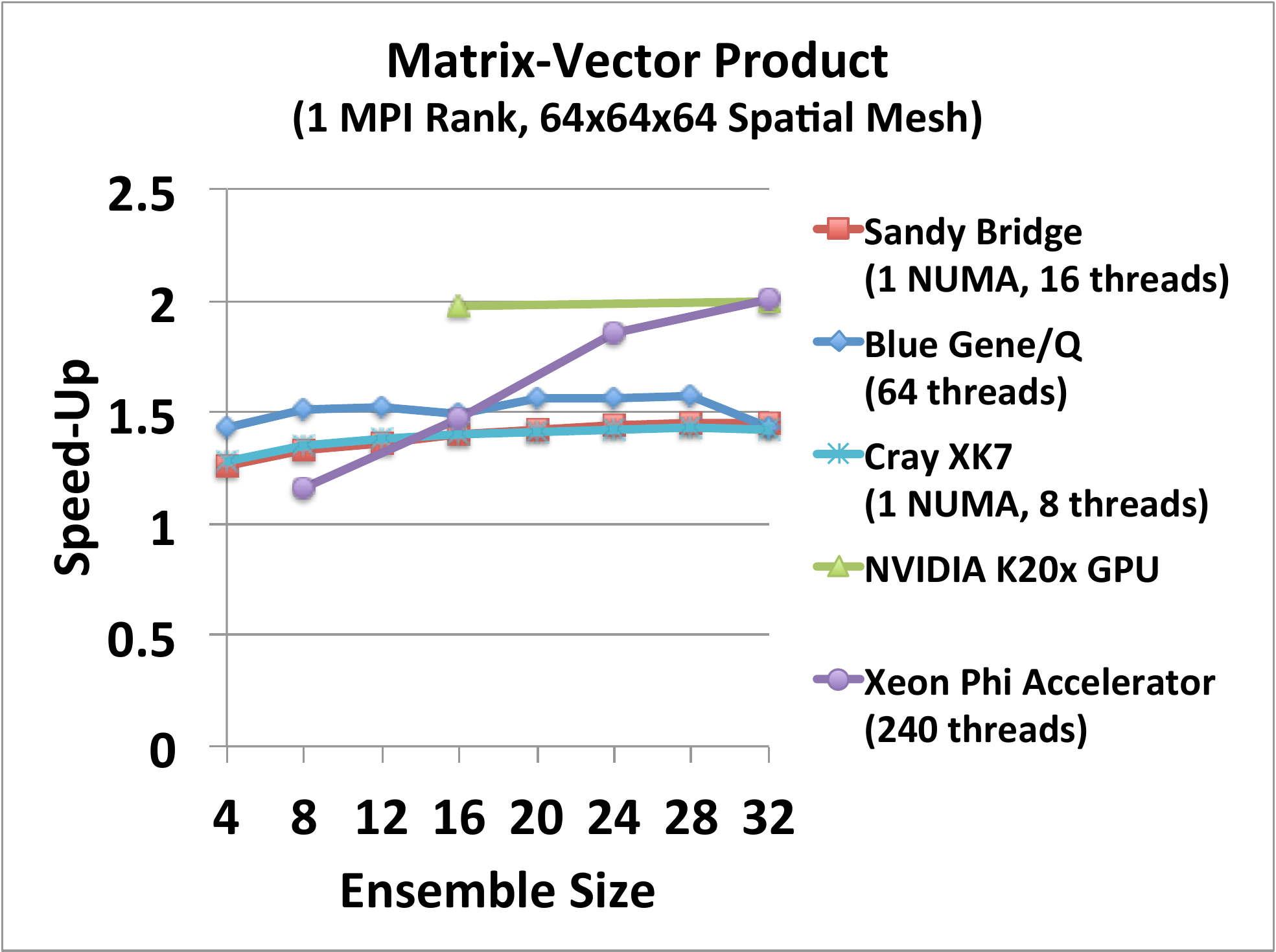}
	\caption{Ensemble matrix-vector product performance.}
	\label{fig:spmv}
\end{figure}

First in \Figref{fig:spmv} we display the measured
speed-up for the ensemble propagation approach applied to the sparse matrix-vector product kernel discussed above for the
matrix $A$ generated on a $64 \times 64 \times 64$ spatial mesh, for a single compute node of each architecture in \Tableref{table:arch}.\footnote{The scalar kernel used in these experiments is a custom matrix-vector product kernel provided by Kokkos that has been optimized for these architectures.  The ensemble kernel is identical to the one described \Figref{fig:ensemble_crs}, but with an additional specialization for CUDA that loads the sparse graph into shared memory to facilitate reuse.}
For reference, we display in \Tableref{table:spmv} the measured sparse matrix-vector product floating-point throughput for each architecture, as well as optimistic and pessimistic bounds for this throughput based on bandwidth considerations.  Assuming storage of 8 B per matrix value, 4 B per matrix column entry, and 8 B per vector value, these bounds are given by
\begin{equation}
\begin{aligned}
	\mbox{Scalar Optimistic} & = \mbox{Bandwidth} \times \frac{\mbox{2 FLOPS}}{\mbox{12 B}} \\
	\mbox{Scalar Pessimistic} & = \mbox{Bandwidth} \times \frac{\mbox{2 FLOPS}}{\mbox{20 B}} \\
	\mbox{Ensemble Optimistic} & = \mbox{Bandwidth} \times \frac{\mbox{2 FLOPS}}{\mbox{8 B}} \\
	\mbox{Ensemble Pessimistic} & = \mbox{Bandwidth} \times \frac{\mbox{2 FLOPS}}{\mbox{16 B}}
\end{aligned}
\end{equation}
where the optimistic bounds assume vector entries are read from a fast cache, and the pessimistic bound assumes they are read from main memory.  The bounds listed in \Tableref{table:spmv} are based on bandwidth measured by the STREAM Triad benchmark~\cite{McCalpin1995}.  For the Sandy Bridge and Cray XK7 architectures with multiple NUMA regions per node, the bandwidth measurements, bounds, and matrix-vector product timings are based on a single NUMA region (8/16 and 8/8 cores/threads respectively).  One can see from \Figref{fig:spmv} that significant improvements in matrix-vector product throughput are achievable by the ensemble approach (even for the relatively well-structured finite element matrix used in this example).  On the CPU architectures, performance of which is better modeled by the optimistic bounds due to their large caches, most of this improvement derives from the 33\% reduction in bandwidth arising from reuse of the matrix graph.  However the speed-ups are larger for the GPU and accelerator architectures where coalesced/packed memory access patterns are required to achieve full bandwidth.
\begin{table*}[t]
  \centering
    \begin{tabulary}{\textwidth}{lCcCCcCC}
    \toprule
    & & \multicolumn{3}{c}{Scalar Mat-Vec} & \multicolumn{3}{c}{Ensemble Mat-Vec} \\
    \cmidrule(lr){3-5} \cmidrule(lr){6-8} \\
    Name & Bandwidth  & Measured & Optimistic & Pessimistic & Measured & Optimistic & Pessimistic \\
         & (GB/s)     &          & Bound      & Bound       &          & Bound      & Bound\\
    \midrule
    Sandy Bridge  & 36.2 & 5.9  & 6.0 & 3.6 & 8.5 & 9.1 & 4.5 \\
    Blue Gene/Q  & 28.5 & 3.5 & 4.8 & 2.9 & 5.0 & 7.1 & 3.6 \\
    Cray XK7       & 11.2 & 2.3 & 1.9 & 1.1 & 3.2 & 2.8 & 1.4 \\
    GPU              & 178 & 17.5 & 29.7 & 17.8 & 34.7 & 44.5 & 22.3 \\
    Accelerator   & 147 & 12.2 & 24.5 & 14.7 & 21.7 & 36.8 & 18.4 \\
    \bottomrule
    \end{tabulary}
    \caption{Measured and expected sparse matrix-vector product throughput (in GFLOP/s) for the scalar and ensemble matrix-vector product (for $s=32$) on each architecture.  Optimistic and pessimistic bounds are based on the main-memory bandwidth as measured by the STREAM Triad benchmark~\cite{McCalpin1995}.}
  \label{table:spmv}
\end{table*}

\begin{figure}[t]
	\centering
	\includegraphics[width=0.45\textwidth]{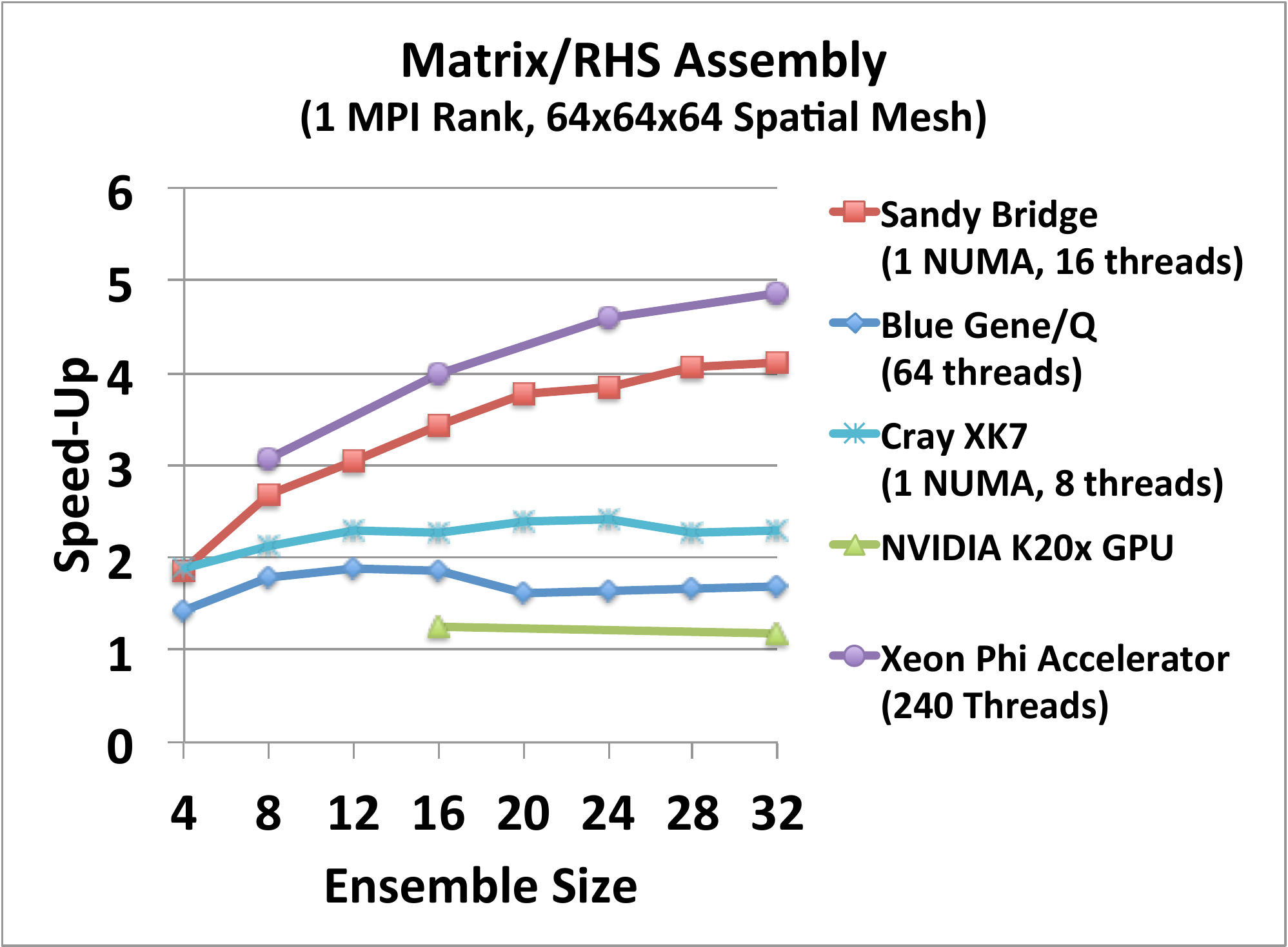}
	\caption{Ensemble linear system assembly performance.}
	\label{fig:assembly}
\end{figure}

We next perform a similar experiment comparing the performance of scalar and ensemble assembly for $A$ and $f$ in
\Eqref{eq:fenl_lin_sys} in \Figref{fig:assembly} (again using only a single NUMA region for the Sandy Bridge and Cray
architectures).  As is clear from \Algreff{alg:assembly} the assembly process involves substantially more operations than the matrix-vector product kernel, with significantly more reuse of data from sample to sample (such as the NodeIndex and ElemGather mappings, and basis function evaluations) and more opportunities for vectorization.  This reuse and vectorization generally results in substantially improved assembly performance for the ensemble approach except for the GPU architecture.  Because of the large operation count, assembly performance is less sensitive to coalesced memory accesses in the sparse gather operation than the matrix-vector product kernel.  Furthermore there is little captured reuse of sample-independent data since reuse can only occur by communicating values between CUDA threads using shared-memory (which for portability reasons is not included in this implementation).

\begin{figure*}
	\centering
	\subfigure[\label{fig:halo_titan_cpu_time}]{
		\includegraphics[width=0.45\textwidth]{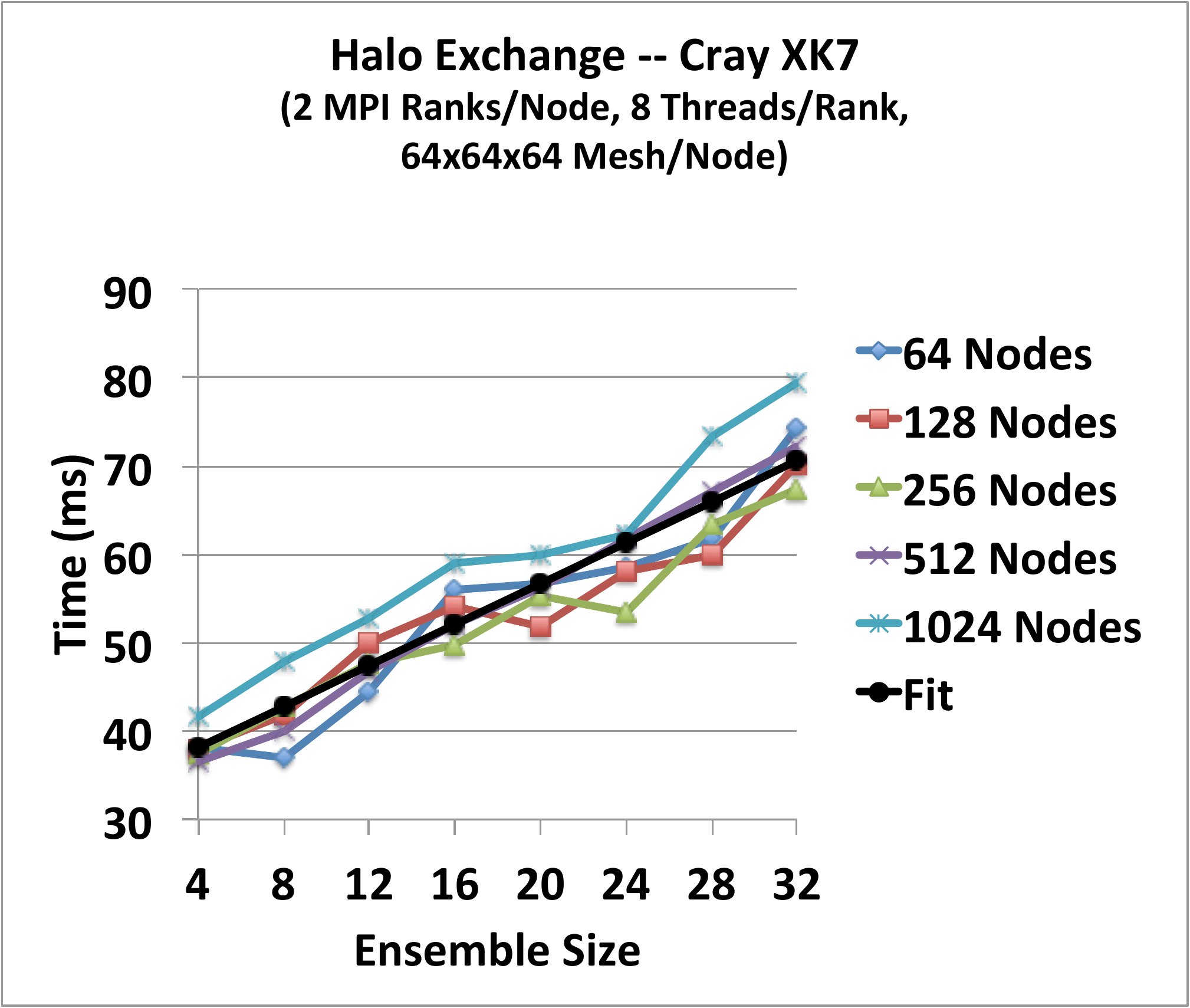}
	}
	\quad
	\subfigure[\label{fig:halo_titan_cpu_speedup}]{
		\includegraphics[width=0.45\textwidth]{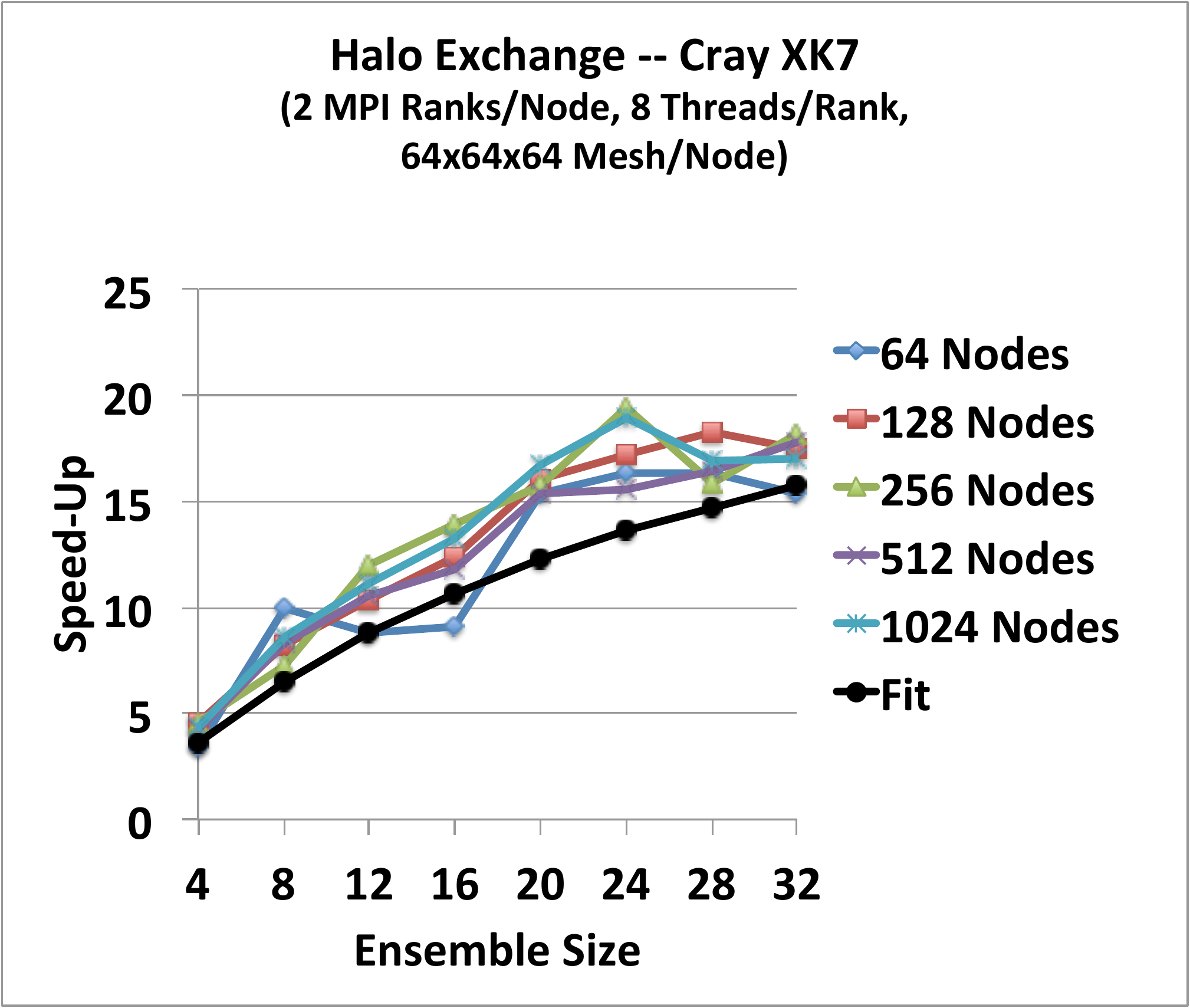}
	}
 	\subfigure[\label{fig:halo_bgq_time}]{
		\includegraphics[width=0.45\textwidth]{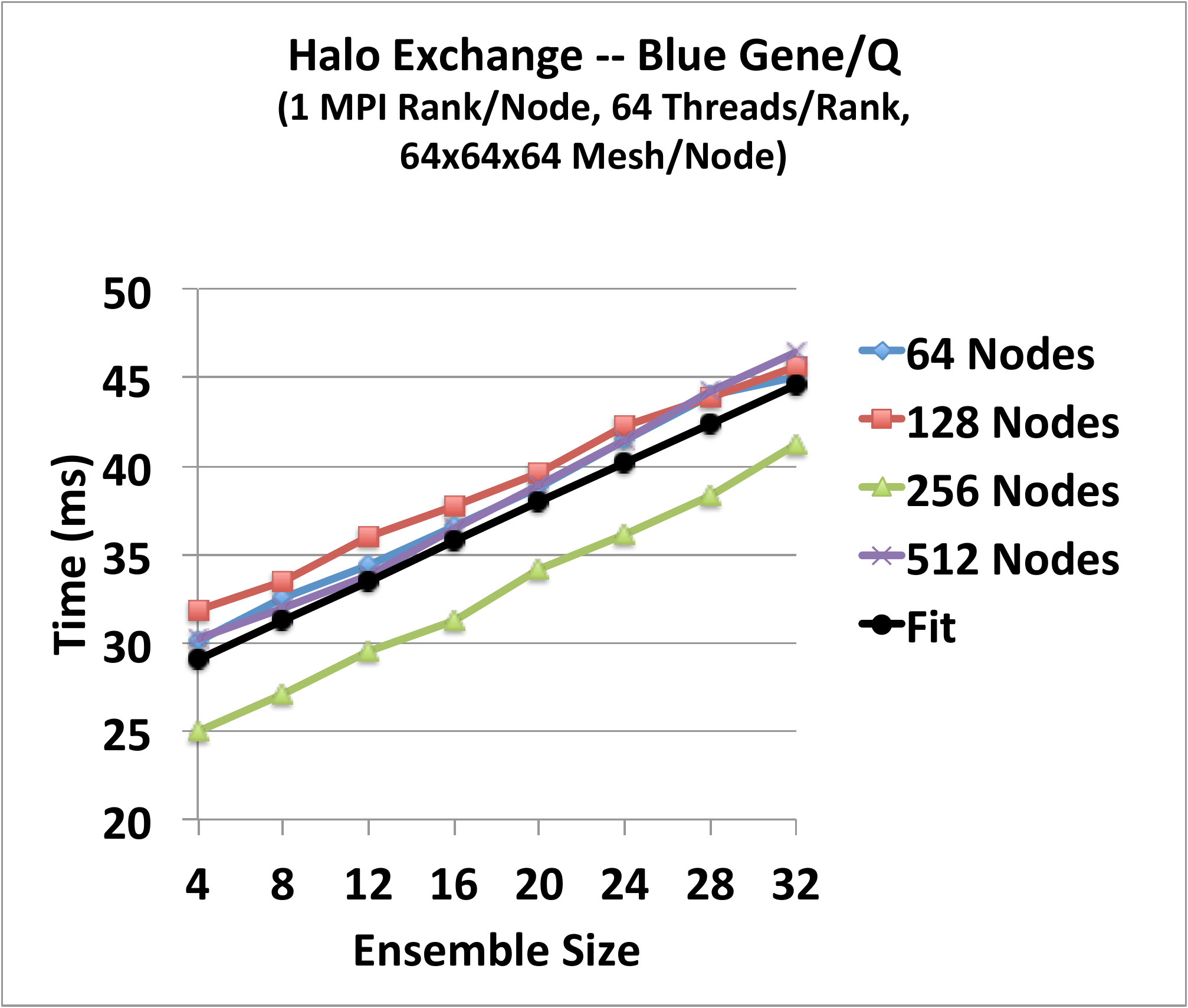}
	}
 	\quad
 	\subfigure[\label{fig:halo_bgq_speedup}]{
		\includegraphics[width=0.45\textwidth]{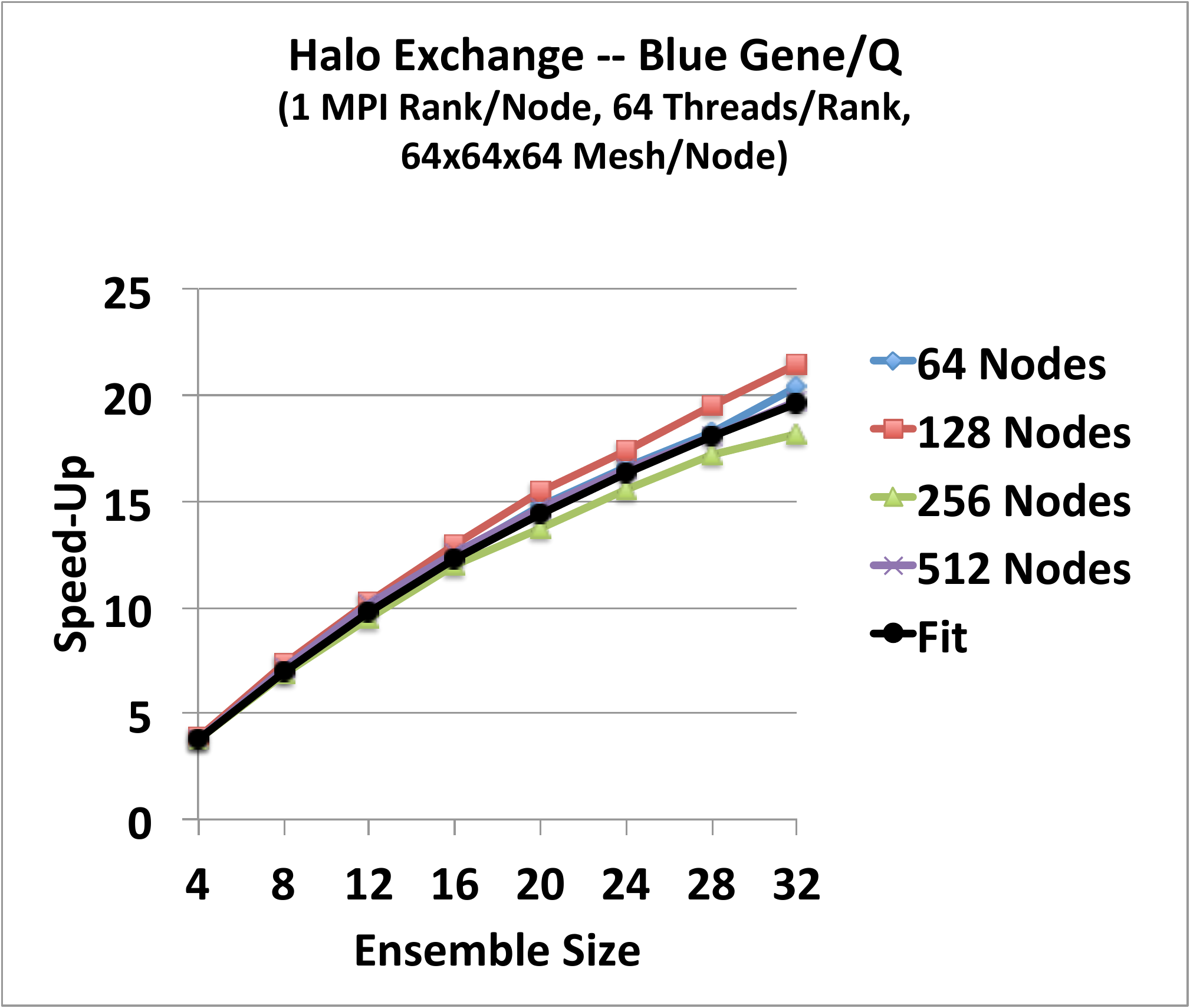}
	}
 	\caption{Ensemble halo exchange performance on Cray XK7 \subref{fig:halo_titan_cpu_time}-\subref{fig:halo_titan_cpu_speedup} and Blue Gene/Q \subref{fig:halo_bgq_time}-\subref{fig:halo_bgq_speedup} architectures, including a linear fit (black curves) of the exchange time (\Eqref{eq:linear_halo_time}) and resulting ensemble speed-up (\Eqref{eq:linear_halo_speedup}).}
 	\label{fig:halo}
\end{figure*}

Next we measure the performance of MPI communication through the halo exchange that is needed for each sparse
matrix-vector product and linear system assembly on both the Cray XK7 and Blue Gene/Q architectures.  In
Figs.~\ref{fig:halo_titan_cpu_time} and~\ref{fig:halo_bgq_time} we display the total ensemble halo exchange time (in
milliseconds) for varying node counts and ensemble sizes, and in Figs.~\ref{fig:halo_titan_cpu_speedup}
and~\ref{fig:halo_bgq_speedup} the corresponding ensemble speed-up.  It is reasonable to assume both halo exchange
message latency and bandwidth are roughly constant with regards to message size over the limited range of ensemble sizes considered, and thus the total exchange time can be approximated by
\begin{equation}\label{eq:linear_halo_time}
	T_{halo}(s) = a + b s,
\end{equation}
where $a$ and $b$ are related to the interconnect latency and bandwidth respectively.
Included in Figs.~\ref{fig:halo_titan_cpu_time} and~\ref{fig:halo_bgq_time} is a plot of \Eqref{eq:linear_halo_time} based on a linear regression for approximating $a$ and $b$ for each architecture (black curve).  Based on \Eqref{eq:linear_halo_time} the speed-up for the ensemble halo exchange is approximately
\begin{equation}\label{eq:linear_halo_speedup}
	\mbox{Speed-up} = \frac{s T_{halo}(1)}{T_{halo}(s)} = \frac{s(a + b)}{a + b s}.
\end{equation}
Also included in Figs.~\ref{fig:halo_titan_cpu_speedup} and~\ref{fig:halo_bgq_speedup} is a plot of \Eqref{eq:linear_halo_speedup} based on the above regression analysis (black curve), which matches the measured data well.  Thus the reduction in total halo exchange time for propagating $s$ samples using the ensemble approach is due to the factor $a s$ reduction in communication latency.

\begin{figure*}
	\centering
	\subfigure[\label{fig:weak_scalar}]{
		\includegraphics[width=0.45\textwidth]{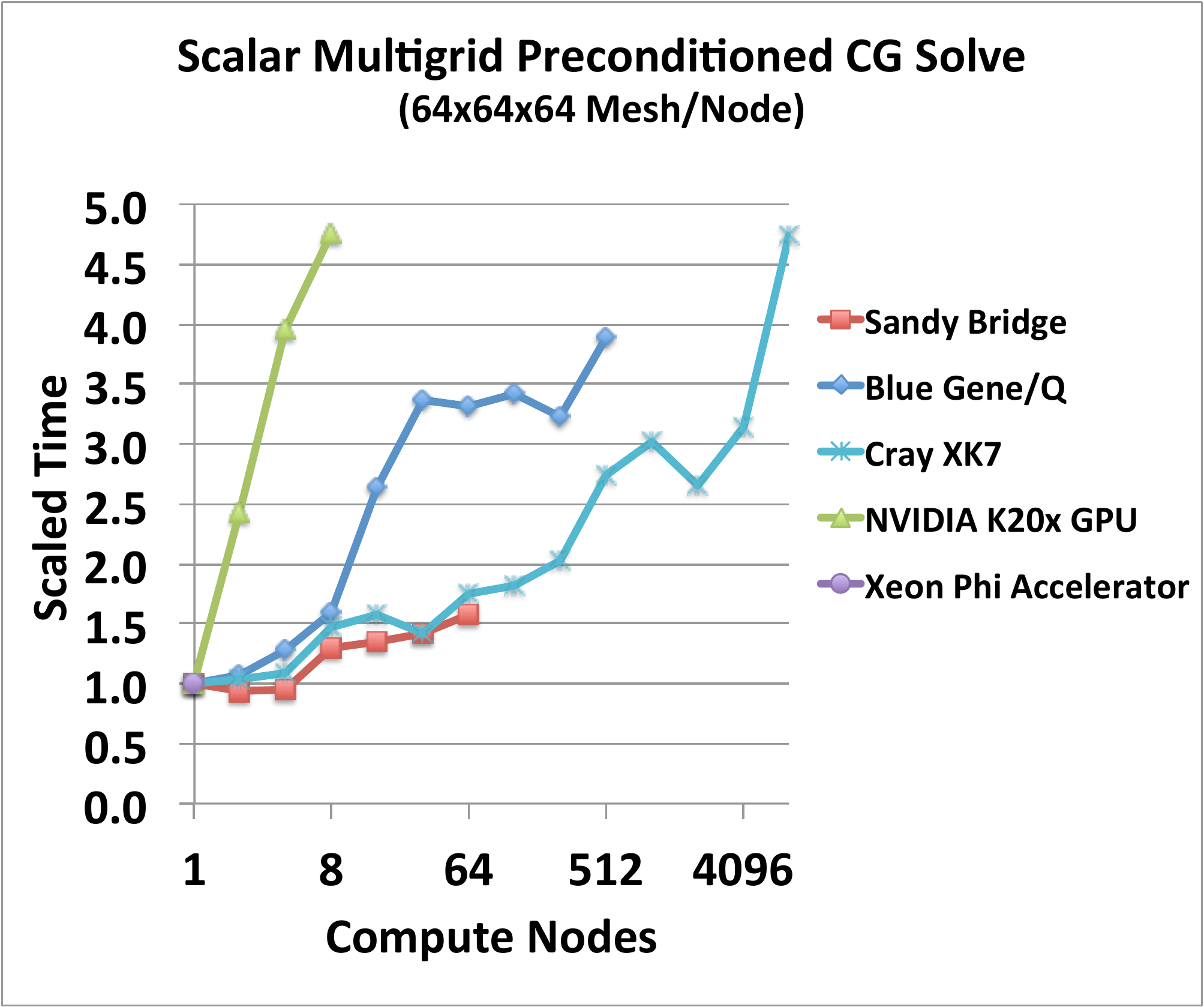}
	}
	\quad
	\subfigure[\label{fig:weak_ensemble}]{
		\includegraphics[width=0.45\textwidth]{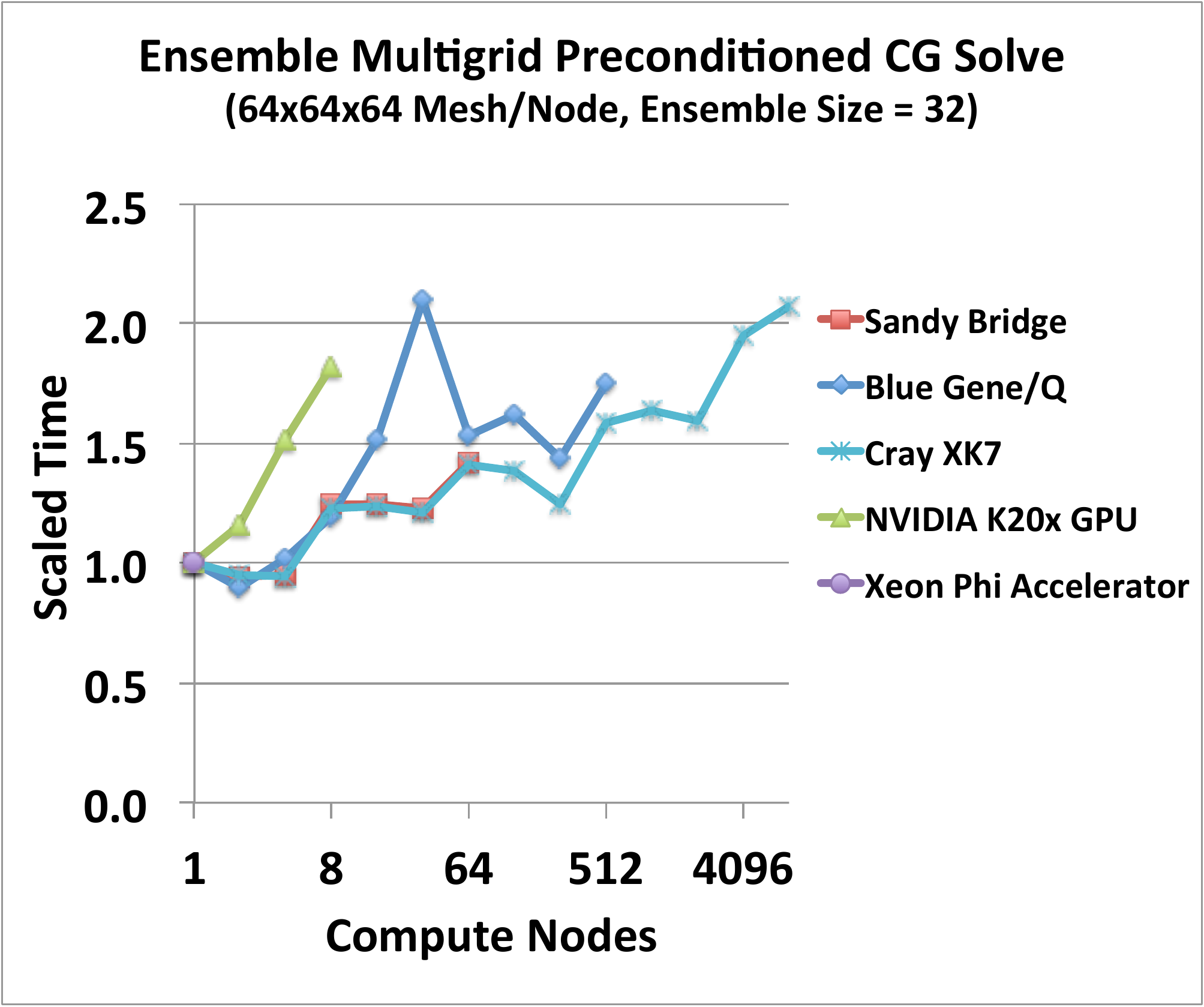}
	}
	\caption{Time for multigrid preconditioned CG solve of scalar system \subref{fig:weak_scalar} and ensemble system \subref{fig:weak_ensemble} relative to time for a single compute node (weak scaling).}
 	\label{fig:weak}
\end{figure*}

\begin{figure}
	\centering
	\includegraphics[width=0.45\textwidth]{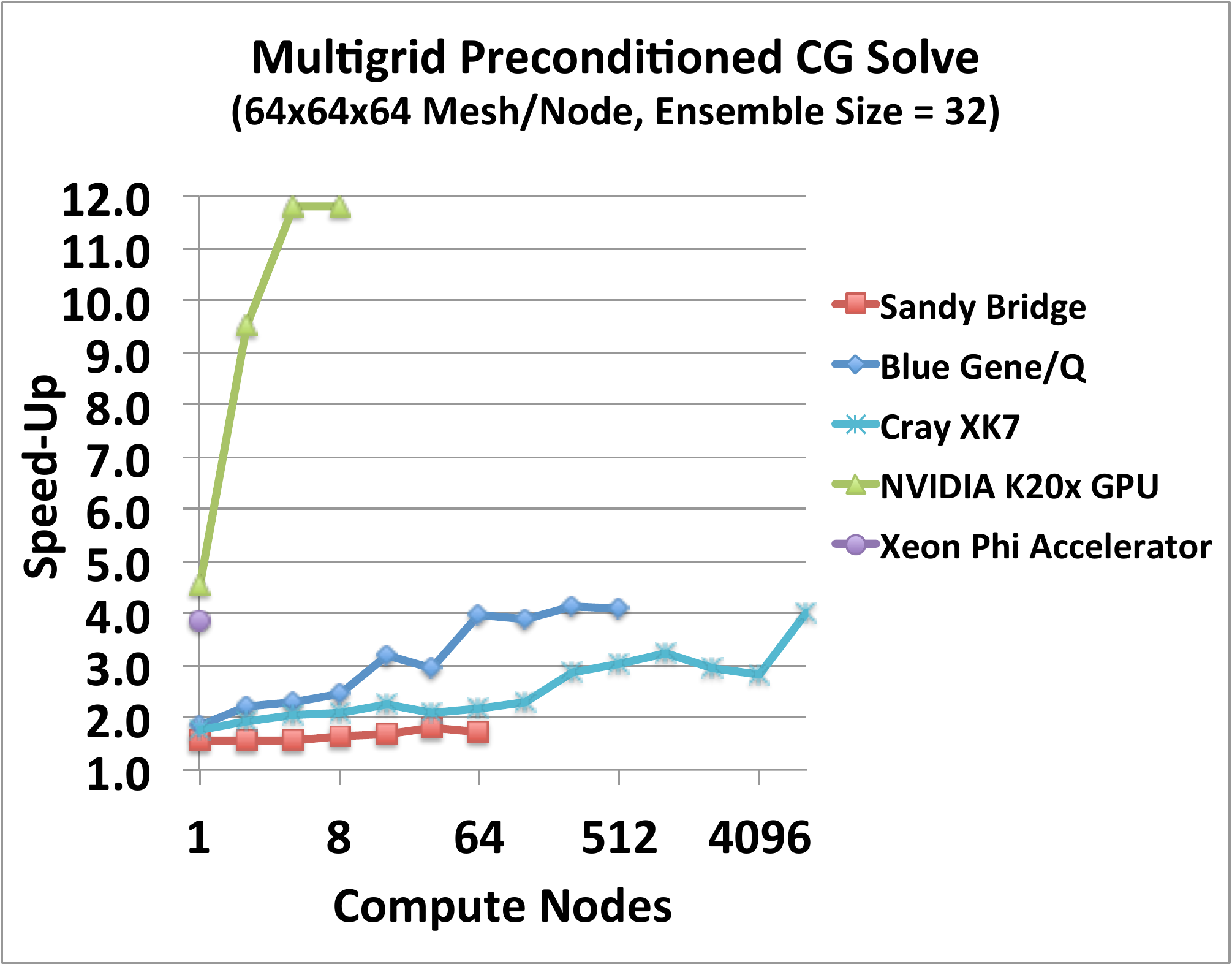}
	\caption{Ensemble preconditioned linear solver speed-up relative to scalar solve.}
	\label{fig:weak_speedup}
\end{figure}

Finally, we investigate the performance of our ensemble approach with multigrid preconditioned CG solves of \Eqref{eq:fenl_lin_sys} for both the Cray XK7 and Blue Gene/Q architectures, as well as 64 nodes of a Sandy Bridge cluster, 8 nodes of a K20X cluster, and a single Xeon Phi 7120P accelerator.  In Figs.~\ref{fig:weak_scalar} and~\ref{fig:weak_ensemble} we show preconditioned CG solve time relative to the time for a single node of each architecture for the scalar and ensemble systems respectively, with the spatial mesh size fixed at $64^3$ cells per compute node as the number of compute nodes is increased (weak scaling).  For the ensemble system, we use $s=32$.  In \Figref{fig:weak_speedup} we display the resulting ensemble solve speed-up. The cost of the preconditioned CG solves is dominated by applications of the multigrid preconditioner, which are in turn dominated by the matrix-vector products within the Chebyshev smoothers and transfer operators.  This leads to improved single-node performance for all architectures, particularly so for the GPU and Accelerator architectures, deriving from the results in \Figref{fig:spmv}.  We also see substantially improved weak-scaling due primarily to the reduced MPI communication costs for the smoothers and transfer operators, particularly for the GPU architecture where communication costs are much higher even at small node counts.


\section{Conclusions}
\label{sec:conclusions}

In this work we described an approach for improving uncertainty propagation performance for sampling-based uncertainty quantification methods based on propagating multiple samples together through scientific simulations.  We argued this should lead to improved aggregate performance by enabling reuse of sample-independent data, improving memory access patterns by replacing sparse gather/scatter instructions with packed/coalesced loads/stores, improving opportunities for fine-grained SIMD/SIMT parallelism, and reducing latency costs associated with message passing.  We described a C++ template and operator overloading approach for incorporating the ensemble propagation approach in general scientific simulation codes, and discussed how tools implementing this technique have been incorporated into a variety of libraries within Trilinos, including the Kokkos manycore performance portability library.  Furthermore, we argued how the approach improves portability by providing uniform access to fine-grained vector parallelism independent of the simulation code's ability to exploit those hardware capabilities.  Finally, we demonstrated performance and scalability improvements for the approach by applying it to the solution of a simple PDE.  

Practical application of these ideas for uncertainty quantification requires grouping samples generated by the uncertainty quantification algorithm into sets of ensembles of a size most appropriate for the architecture.  In the experiments covered in this paper, we found an ensemble size of 32 generally works well for all architectures considered.  Grouping samples appropriately is critical for the approach to be effective for real scientific and engineering problems, where the simulation process should be as similar as possible for all samples within an ensemble.  For example, the spectra of matrices used in the solution of linear/nonlinear systems should be similar to prevent growth in solver iterations.  Furthermore the code-paths required for samples within an ensemble need to be similar to achieve effective speed-ups.  Work exploring algorithmic grouping approaches is underway and will be discussed in a subsequent publication.  However once samples have been grouped, each ensemble can be propagated independently using the traditional coarse-grained distributed memory approach.

\bibliographystyle{siam}
\bibliography{paper}  

\end{document}